\documentclass[%
twocolumn
,reprint
,longbibliography
,notitlepage
,floatfix
,balancelastpage
,superscriptaddress
,amsmath
,amssymb
,aps
,pra
]{revtex4-1}

\usepackage{graphicx}
\usepackage{dcolumn}
\usepackage{bm}
\usepackage{braket}
\usepackage{amsthm}
\usepackage{subfigure}
\usepackage{url}
\usepackage[normalem]{ulem}
\usepackage[usenames,dvipsnames]{xcolor}
\usepackage[pdftex,letterpaper=true,pagebackref=false]{hyperref}
\usepackage[nameinlink,capitalize]{cleveref}

\newcommand{\dket}[1]{\mbox{$\left|\!\left.#1\right\rangle\!\right\rangle$}}

\newcommand{\dbradket}[2]{\mbox{$\langle\!\langle #1|#2\rangle\!\rangle$}}
\newcommand{\dketdbra}[2]{\mbox{$|#1\rangle\!\rangle\!\langle\!\langle #2|$}}

\hypersetup{
pdftitle={Effect of error and its mitigation in quantum circuit cutting},
pdfauthor={Ritajit Majumdar and Christopher J. Wood},
	pdfnewwindow=true, 
	colorlinks=true, 
	linkcolor=Black,
	citecolor=Magenta,
	urlcolor=MidnightBlue
}

\begin{document}

\title{Error mitigated quantum circuit cutting}

\author{Ritajit Majumdar}
\email{majumdar.ritajit@gmail.com}
\affiliation{Advanced Computing \& Microelectronics Unit, Indian Statistical Institute, India}

\author{Christopher J. Wood}
\email{cjwood@us.ibm.com}
\affiliation{IBM Quantum, IBM T.J. Watson Research Center, Yorktown Heights, NY 10598, USA}

\date{November 24, 2022}

\begin{abstract}

We investigate an error mitigated tomographic approach to the quantum circuit cutting problem in the presence of gate and measurement noise. We explore two tomography specific error mitigation techniques; readout error mitigated conditional fragment tomography, which uses knowledge of readout errors on all cut and conditional qubit measurements in the tomography reconstruction procedure; and dominant eigenvalue truncation (DEVT), which aims to improve the performance of circuit cutting by performing truncation of the individual conditional tomography fragments used in the reconstruction. We find that the performance of both readout error mitigated tomography and DEVT tomography are comparable for circuit cutting in the presence of symmetric measurement errors. For gate errors our numerical results show that probability estimates for the original circuit obtained using DEVT outperforms general circuit cutting for measurement, depolarization and weakly biased Pauli noise models, but does not improve performance for amplitude damping and coherent errors, and can greatly decrease performance for highly biased Pauli noise. In cases where DEVT was effective, it as also found to improve performance of partial tomographic reconstruction using at least 50\% of the full tomographic data with a conditional least-squares tomographic fitter, while linear inversion tomography with or without DEVT mitigation was found to perform poorly with with partial data.
\end{abstract}

\maketitle

\section{Introduction}
\label{sec:intro}

The limited number of qubits on near-term quantum devices is a significant limitation to the size and type of quantum computation problems that can be evaluated on them. Circuit knitting, an umbrella term for combing results of two smaller quantum processors to logically form a larger device, has been suggested as a path to scalability~\cite{bravyi2022aip}. For such a logical device, it is useful to \emph{cut} a bigger quantum circuit into smaller pieces so that each of them can be executed on the hardware at hand.  Cutting can be broadly classified into (i) cutting the wire (termed as circuit cutting henceforth) \cite{peng2020simulating, perlin2021quantum, tang2021cutqc, lowe2022fast, uchehara2022rotation}, (ii) replacing two qubit gates by mid circuit measurements and classical feed-forward conditional options (often called gate cutting) \cite{mitarai2021constructing, piveteau2022circuit, mitarai2021overhead}, and (iii) partitioning the problem into multiple weakly interacting sub-problems (e.g. entanglement forging \cite{eddins2022doubling}, frozen-qubit QAOA \cite{ayanzadeh2022frozenqubits}). In this work we focus on the study of circuit cutting.

\begin{figure*}[htb]
    \centering
    \subfigure[An example circuit]{
        \includegraphics[scale=0.16]{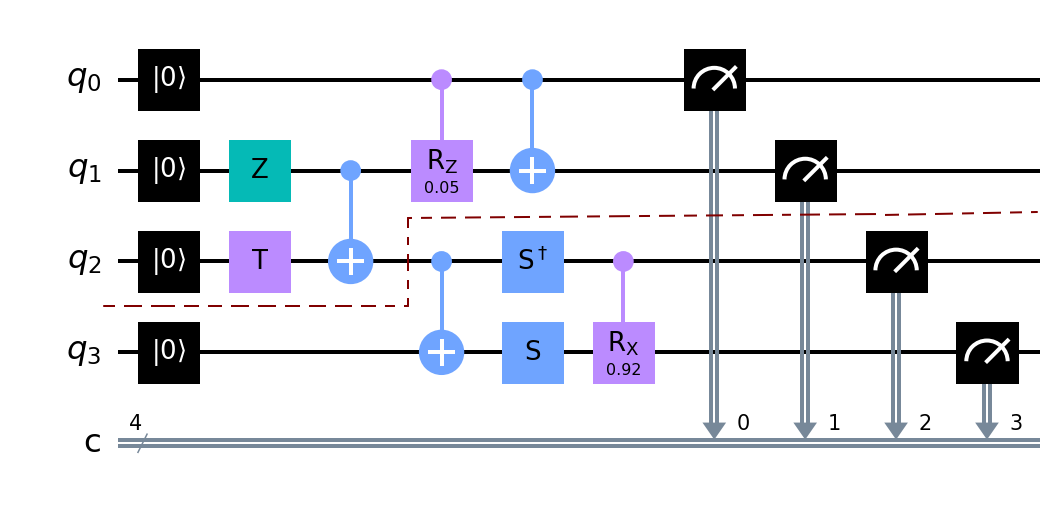}
        \label{fig:ex_circ_a}
    }
    \subfigure[First subcircuit]{
        \includegraphics[scale=0.19]{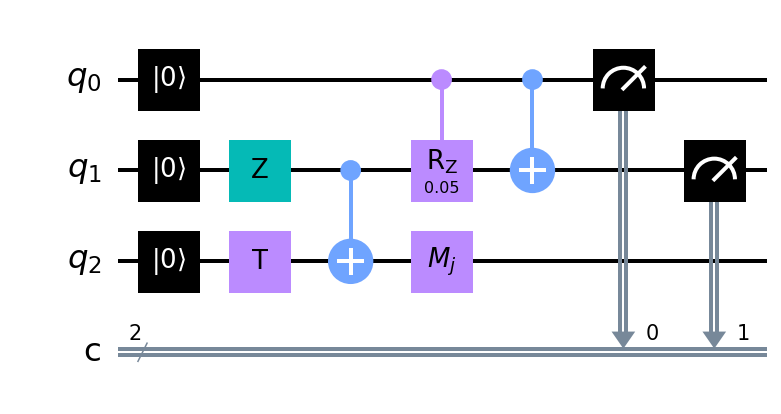}
        \label{fig:ex_circ_b}
    }
    \subfigure[Second subcircuit]{
        \includegraphics[scale=0.22]{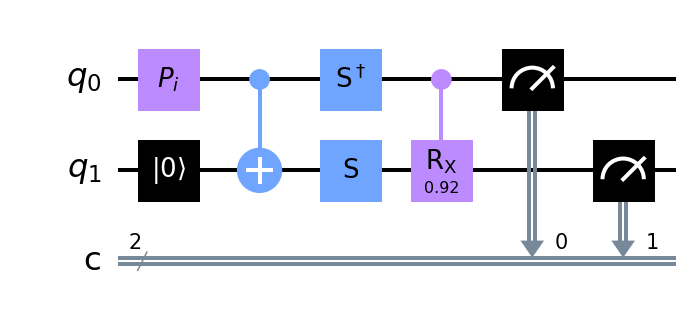}
        \label{fig:ex_circ_c}
    }
    \caption{An example of cutting of a random 4 qubits quantum circuit into 2 fragments. $P_i$ and $M_j$ denotes tomographically complete preparation and measurement basis respectively. The red dotted line on Fig. (a) shows the cut location.}
    \label{fig:ex_circ}
\end{figure*}

In circuit cutting, the quantum circuit is \emph{cut} into two or more \emph{fragments} such that each fragment is small enough to be computed on the quantum hardware individually. Note that here the cutting is along the wire only, and all the gates from the original circuit is retained in the ensemble of the fragments (refer to \cref{fig:ex_circ}). The expectation value of the original circuit can be retrieved by classically combining the expectation values of the individual fragments obtained in different preparation and measurement bases \cite{peng2020simulating}. The post-processing time, however, scales exponentially with the number of cuts, thus making a large number of cuts infeasible. Since reconstruction of expectation values of the full circuit assumes accurate estimation of expectation values of each fragment, statistical errors due to finite samples of each fragment can lead to an invalid distribution which may not be non-negative, nor correctly normalized. By using maximum-likelihood tomography to constrain each fragment to a valid physical state or channel, as was proposed in \cite{perlin2021quantum}, this will result in a valid final distribution.

Circuit cutting presents several opportunities for error mitigation in addition to those that can be applied to standard circuits. Since individual fragments contain fewer gates than the original circuit, as can be seen in the example in \cref{fig:ex_circ}, individual fragments may contain less overall noise~\cite{ayral2021quantum, basu2021qer}, which may make them more amenable to error mitigation techniques such as probabilistic error cancellation which exhibit exponential scaling with the total noise strength of the circuit \cite{berg2022probabilistic}. Furthermore, since tomography is used as subroutine for reconstruction we can apply mitigation techniques during the fitting which would not be possible otherwise, such as eigenvalue truncation or re-scaling. This is notably different to typical tomography applications for device characterization, where the goal is to get as accurate an estimate of noisy states and channels as possible. In this paper, we focus on classical measurement readout errors, and a variety of representative gate error channels including depolarization, pauli, amplitude damping, and coherent rotation errors, on a type of random trotterized circuit (\cref{fig:cluster_unitary_f2q4}), also used in \cite{perlin2021quantum}, which finds applications in variational quantum algorithms \cite{cerezo2021variational}. We study the effect of each of these noise on circuit cutting, and the improvements obtained by applying tomography specific error mitigation techniques, namely measurement error mitigation (MEM) \cite{nation2021scalable} and dominant eigenvalue truncation (DEVT) \cite{koczor2021dominant}, to the fragments used in the reconstruction of the circuit outcome distribution.

In this numerical study, we first explore the applicability of MEM on tomography, and show that it increases the fidelity of tomographic quantum circuit cutting. In the presence of gate errors, which is often much more difficult to deal with than measurement error, MEM is not sufficient. We show that DEVT offers significant improvement over general circuit cutting for depolarization and pauli noise models. Moreover, DEVT alone is shown to be sufficient to account for measurement error, thus obliterating the requirement of MEM. Although the performance of DEVT deteriorates for amplitude damping and coherent noise models, it still provides improvement over circuit cutting without mitigation. Both the types of tomographic fitters considered in this paper, namely Conditional Least Square (CLS) and Linear Inversion (LIN), are shown to perform equivalently when equipped with DEVT. Finally, CLS is shown to retain a near-optimal performance for reconstruction using partial tomographic data when at least $50\%$ of the tomographic data is used, whereas LIN performs poorly with partial data both in the presence or absence of DEVT.

The rest of the paper is organized as follows. We introduce tomographic circuit cutting and conditional fragment tomography in \cref{sec:circuit_cutting} and discuss about its scalability. In \cref{sec:mit} we introduce tomographic error mitigation techniques for  measurement error mitigated conditional fragment tomography and dominant eigenvalue truncation (DEVT) to counter measurement noise in the system. \cref{sec:simulations} shows the simulation details of the effect of noise and error mitigation on quantum circuit cutting, and also compares two tomographic approaches for circuit cutting with partial data. The analysis of these noise models and their effects are contained in the appendices.

\section{Tomographic Circuit Cutting}
\label{sec:circuit_cutting}

Circuit cutting has been introduced as a procedure for estimating a measurement output distribution or expectation value of a quantum circuit on $N$ qubits by post-processing the outcomes of a set of measurements performed on a collection of circuit \emph{fragments} each using fewer qubits than the original circuit~\cite{peng2020simulating}. Given a circuit with $N$ qubits, where $N$ may be larger than the number of qubits on the hardware, circuit cutting involves separating the circuit into a set of smaller disconnected fragments $\{\mathcal{F}_\alpha\}$ by \emph{cutting} qubit-wires at specific locations. This is illustrated in \cref{fig:ex_circ} where \cref{fig:ex_circ_a} is an example circuit to be cut into two fragments which are shown in \cref{fig:ex_circ_b} and \cref{fig:ex_circ_c}. The choice and number of cut locations depends on the circuit topology and it is an active area of research to develop application circuits and cutting algorithms amenable to efficient circuit cutting \cite{tang2021cutqc, basu2021qer}.

\begin{figure*}
    \centering
    \subfigure[1-qubit quantum state fragment]{
        \includegraphics[scale=0.28]{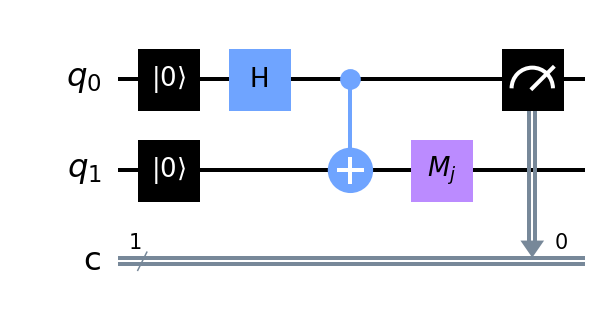}
    }
    \subfigure[1-qubit quantum channel fragment]{
        \includegraphics[scale=0.28]{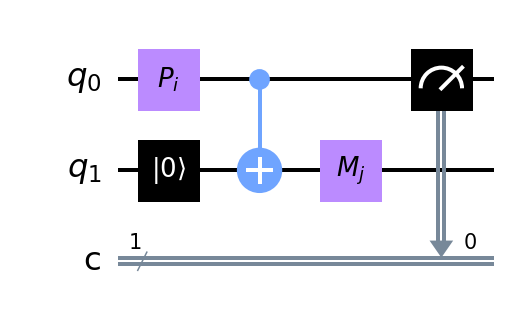}
    }
    \subfigure[1-qubit POVM fragment]{
        \includegraphics[scale=0.28]{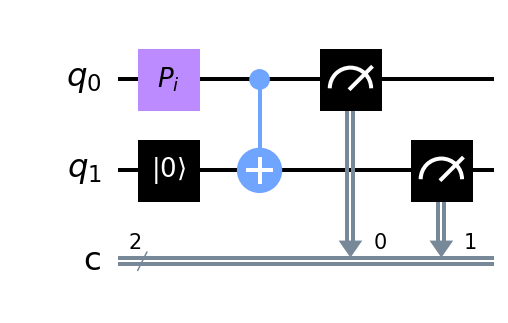}
    }
    \caption{An example cutting of a 4 qubits GHZ state into 3 fragments. A fragment with only a single tomographic measurement $M_j$ behaves as a quantum state fragment, one with only a single tomographic preparation $P_i$ behaves as a POVM fragment, whereas a general fragment with both tomographic preparation and measurement is a quantum channel fragment. Note that the number of effective qubits of the conditional tensor in the fragment corresponds to the number of tomographically prepared or measured qubits, which is 1-qubit in all shown cases, rather than the total number of qubits in the fragment.}
    \label{fig:frag_types}
\end{figure*}

\subsection{Fragment Reconstruction}

The general goal of circuit cutting is to reconstruct either the full probability distribution $P(s)$ for all measurement outcomes of the original circuit, or some quantity derived from them such as an expectation value. Each fragment $\mathcal{F}_\alpha$ of a cut circuit will include some non-overlapping subset of the original measurements along with 1 or more unconnected qubit-wires at each cut location. As shown in \cref{fig:frag_types}, these fragments will be equivalent  to either a quantum state, a quantum channel, or a POVM depending on whether the locations of the cut wires contain only open output wires, both input and output wires, or only input wires respectively. 

Let $m_\alpha$ be the number of original circuit measurement in $\mathcal{F}_\alpha$, the outcomes of which we will index by $s_\alpha \in \{0, 1\}^{m_\alpha}$, and let $k_\alpha$ be the number of qubit wires in the fragment associated with a cut. In this case the fragment can be represented as a block-diagonal $n_\alpha = k_\alpha + m_\alpha$ qubit tensor
\begin{align}
    T_{\mathcal{F}_\alpha} &= \sum_{s \in \{0,1\}^{m_\alpha}}
        T_\alpha(s_\alpha) \otimes \ket{s_\alpha}\!\!\bra{s_\alpha}
        \label{eq:cond-tensors}
\end{align}
where $T_\alpha(s_\alpha)$ is the $k_\alpha$-qubit tensor component representing the unnormalized density matrix, Choi-matrix, or POVM element, corresponding to the measurement outcome $s_\alpha$ on the $m_\alpha$-qubit measurements. The probability of outcome $s_\alpha$ is given by
\begin{align}
    p_\alpha(s_\alpha) = \frac{Tr[T_\alpha(s_\alpha)]}{Tr[T_{\mathcal{F}_\alpha}]}.
\end{align}

The original circuits measurement output distribution $P(s)$ can be reconstructed from the fragments in \cref{eq:cond-tensors} by a series of independent tensor contractions for each outcome probability
\begin{align}
P(s) &= C(T_1(s_1), \hdots, T_f(s_f))
\label{eq:circ-probs}
\end{align}
where $s$ is the union of fragment outcomes $\{s_1, ..., s_f\}$ and $C(T_1(s_1), \hdots, T_f(s_f))$ is the tensor contraction of all the conditional fragment components. For example, for three fragments where one is a state, two is a channel, and three is a POVM this is given by
\begin{align*}
    C(\rho_1(s_1), \mathcal{E}_2(s_s), \Pi_3(s_3))
     = \mbox{Tr}[\Pi_3(s_3)\cdot\mathcal{E}_2(s_2)(\rho_1(s_1))]
\end{align*}
In the remainder of the manuscript when talking about a single fragment we will drop the fragment index $\alpha$ to simplify the notation.

As shown in \cite{perlin2021quantum} circuit cutting can be considered as a composite tomographic problem where we estimate each of the conditional tensor components $T_\alpha(s_i)$ in \cref{eq:cond-tensors} via an appropriate state, process, or measurement tomography experiment, and then reconstruct the full probability distribution $P(s)$ of the original circuit via the tensor contractions in \cref{eq:circ-probs}.

\subsection{Quantum Tomography}\label{sec:tomo}

For this work we consider a general description of tomography of a tensor $T$. This encompases state tomography when $T=\rho$ corresponds to a density matrix, process tomography when $T=\Lambda$ is a Choi-matrix, and measurement tomography when $T=M_j$ is a POVM element. In all cases quantum tomography of $T$ consists of choosing a basis $\{B_j\}$ of tensors that spans $T$, where such a spanning set is called \emph{tomographically complete}, that we can use to experimentally measure the set of measurement probabilities $p_j = \dbradket{B_j}{T}$, where $\dket{T}$ denotes \emph{vectorization} of the tensor $T$~\cite{gilchrist2009arx}. For state, process, and measurement tomography this basis can be chosen as $B_i = M_i$, $B_{ij} = \rho_i^T\otimes M_j$, $B_i=\rho_i^T$ respectively, where $\{\rho_i\}$ is a tomographically complete preparation basis of input states, and $\{M_j\}$ is a tomographically complete basis of measurement POVMs. 

If $\{B_j\}$ is tomographically complete, then the probabilities $\{p_j\}$ contain sufficient information to completely reconstruct $T$. In this work we consider two reconstruction methods which both implement a form of maximum-likelihood estimation. The first is \emph{linear inversion} combined with re-scaling of the fitted tensor to enforce positivity as described in~\cite{smolin2012efficient}, and the second is constrained linear-least squares estimation implemented as \emph{semidefinite program} optimization problem.

\subsubsection{Linear Inversion}\label{sec:linear_inversion}

For a  tomographically complete basis $\{B_j\}$ and outcome probabilities $\{p_j\}$ linear inversion amounts to constructing a \emph{dual basis}~\cite{dariano2000pla} $\{D_j\}$ defined by the orthogonality relation $\dbradket{D_i}{B_j}=\delta_{ij}$ as
\begin{align}
    \dket{D_j} = \left(\sum_i \dketdbra{B_i}{B_i}\right)^{-1}\dket{B_j}
    \label{eq:dual-basis}
\end{align}
The linear inversion estimate of the tensor $T$ is then given by
\begin{align}
    T_{LIN} = \sum_i p_i D_i.
\end{align}
The linear inversion estimate of a state or channel is generally not positive or completely-positive respectively, however by performing a specific re-scaling of eigenvalues this will result in a physical state that is consistent with the maximum likelihood estimated value under the assumption of Gaussian measurement noise~\cite{smolin2012efficient}.

\subsubsection{Constrained Least-Squares}\label{sec:lstsq}

For a  tomographically complete basis $\{B_j\}$ and outcome probabilities $\{p_j\}$ constrained least-squares tomography is the optimization problem given as
\begin{align}
    T_{LS} = \mbox{arg}\min_{T\ge 0} \frac12\left\|\Sigma^{-1/2}\left(
    S_B\dket{T} - \ket{p}
    \right)\right\|_2^2
    \label{eq:lstsq}
\end{align}
where $\ket{p} = \sum_i p_i \ket{i}$ is a vector of measured outcome probabilities, $\Sigma^{-1/2}$ is a covariance matrix for the measurement outcome probabilities $\{p_j\}$ and $S_B\dket{T} = \sum_i \ket{i}\!\!\dbradket{B_i}{T}$ is a vector of expected probabilities for the model $T$.

Typically we also include additional constraints in \cref{eq:lstsq}, such as $T$ is trace 1 for state tomography, trace-preserving for process tomography, or the sum of POVM elements is the identity for measurement tomography. In all these cases these constraints are positive-semidefinite and the resulting optimization problem is a semidefinite program.

\subsection{Conditional Fragment Tomography}\label{sec:frag-tomo}

We now generalize the description of tomography from the previous subsection to apply to circuit cutting fragments as a form of conditional tomography. Consider a cut fragment $\mathcal{F}$ with $m$-qubit measurements corresponding to the original circuits outputs, and $k$ qubits corresponding to the cut qubits as either a state, channel, or POVM fragment.

Rather than try and reconstruct a full description of the fragment tensor in \cref{eq:cond-tensors} which would be inefficient, we take advantage of the block-diagonal structure and instead reconstruct the set of conditional fragment components $\{T(s)\}$ for $s\in \{0, 1\}^m$. This is done by choosing a tomographically complete basis $\{B_j\}$ on the $k$ cut qubit subsystem, and also defining an orthonormal (but tomographically incomplete) basis $\{\Pi_{s}\}$ on the $m$ conditional measurement outcome qubits. Typically this second basis is chosen as computational basis $\Pi_{s} = \ket{s}\!\!\bra{s}$. With these two basis we can consider tomography of the fragment tensor $T_{\mathcal{F}}$ in terms of the tensor product basis $B_i\otimes \Pi_s$, where the probability of observing an outcome $(i,s)$ is given by
\begin{align}
    p_{i,s} &= \dbradket{B_i\otimes \Pi_s}{T_{\mathcal{F}}}
        = \sum_{s^\prime} \dbradket{B_i}{T(s^\prime)} \bra{s^\prime}\Pi_s\ket{s^\prime}
    \label{eq:frag-tomo-probs}
\end{align}
which for $\Pi_{s} = \ket{s}\!\!\bra{s}$ gives
\begin{align}
    p_{i,s} &= \dbradket{B_i}{T(s)}.
    \label{eq:frag-tomo-probs-diag}
\end{align}

Using this we can perform tomographic reconstruction of $T(s)$ via linear inversion, least-squares optimization, or any other tomography fitter by fixing the index $s$ and performing tomography fitting using the set of probabilities $\{p_{i, s}\}$. This means that in general for a fragment $\mathcal{F}$ with $m$ conditioning measurements, and $k$ cut qubits there are $2^m$ conditional $k$-qubit tensor tomography fitting procedures to run on the measurement data.

\subsection{Scalability Considerations}

For a general $k$-qubit quantum channel fragment measured using a 4-element preparation basis, and the 6 outcomes of Pauli measurements in the X, Y, Z bases the number of tomographic circuits for reconstructing a fragment is $12^k$. This limits the number of cut qubits per fragment that is feasible in applications. Furthermore, even after reconstructing all conditional fragment components for each fragment, there are $2^n$ tensor contractions that must be performed to reconstruct a full probability distribution of an $n$-qubit uncut circuit. However, there are myriads of problems in quantum chemistry, combinatorial optimization, quantum machine learning etc. that only require computing the expectation value of low-weight Pauli observables, not the full distribution, and circuit cutting can be implemented more efficiently in these cases to reduce the exponential number of tensor contractions to a polynomial number.

In order to evaluate the expectation value of a weight $d < m$ Pauli operator for a fragment $\mathcal{F}$ with $k$ cut qubits and $m$ qubit measurements from the original circuit, it suffices to evaluate the conditional components $T_s$ only over the $d$ non-identity qubits corresponding to the observable, and marginalize over the rest. This reduces the complexity tensor contractions required to compute $\langle P\rangle$ to 
$\mathcal{O}(2^d \cdot \begin{pmatrix}m\\d\end{pmatrix})$.
For a fixed $d$, $2^d$ is $\mathcal{O}(1)$, and $\begin{pmatrix}m\\d \end{pmatrix}$ is $\mathcal{O}(m^d)$, which becomes the effective complexity of tensor contraction.

To take a more concrete example, consider estimation of a Hamiltonian
$$H = \sum_i c_i Z_i + \sum_{i \neq j} \chi_{ij}Z_i Z_j$$
which requires only one and two body interactions over the qubits. Such Hamiltonians are extremely common in Quantum Chemistry, and other hardware-efficient near-term applications. For such a Hamiltonian, there will be $\begin{pmatrix}
m\\
2
\end{pmatrix}$ weight-2 observables, and $m$ weight-1 observables per fragment. Therefore, the over-all complexity of tensor contraction boils down to $\{2^2 \cdot \mathcal{O}(m^2) + 2 \cdot \mathcal{O}(m)\}$ which is $\mathcal{O}(m^2)$. For most practical purpose, $m$ scales as $\mathcal{O}(n)$ where $n$ is the total number of qubits in the fragment. For example, when there are $k$ cut qubits in a fragment, $m = n-k$. Therefore, the complexity of tensor contraction can be lowered to $\mathcal{O}(n^2)$ if finding expectation values of weight-2 Pauli observables are sufficient.

For the numerical simulations in this paper we shall still consider the full probability distribution of the original uncut circuit, as was done in previous works \cite{peng2020simulating, perlin2021quantum, tang2021cutqc}, since this is a more stringent test of the quality of the reconstruction.

\section{Error Mitigation}\label{sec:mit}

In this work we propose and investigate the performance of two new forms of tomography specific error mitigation which are not possible with standard circuit execution. These are (i) \emph{readout error mitigated conditional tomography}, which aims to remove the effect of measurement errors during the tomographic reconstruction by performing a constrained fit of all conditional fragments simultaneously using the knowledge of the readout error model, and (ii) \emph{dominant eigenvalue truncation} (DEVT) which involves truncation of the reconstructed state or channel to its largest eigenstates.

\subsection{Measurement Error Mitigated Conditional Tomography}\label{sec:meas-mit-tomo}

A significant source of error in tomographic experiments are so called state preparation and measurement (SPAM) errors. If an ideal preparation and measurement basis is used in a tomography fitter, any errors in these processes will be attributed to errors in the reconstructed state or channel itself. In currently available quantum devices measurement errors are the dominant source of SPAM error and are typically in the range of 0.5\%-10\% for a single-qubit measurement depending on the architecture~\cite{nation2021scalable}. Due to this a variety of error mitigation schemes have been proposed for mitigating classical readout errors which involve some characterization process of the measurement error model, and processing of measurement outcomes to attempt to undo these effects \cite{cai2022arx}. When performing tomography it is inadvisable to apply such mitigation techniques to process counts before using them during tomographic fitting. Instead, one can perform general measurement error mitigation as part of the fitting procedure if they have a well characterized measurement error model by using the noisy POVM elements directly in the tomography fitter basis, which can then be used to construct a noisy error mitigated dual basis via \cref{eq:dual-basis} for linear inversion, or used directly in the least-squares objective function in \cref{eq:lstsq}.

Conditional tomography presents a challenge since this kind of error mitigation can only be applied to the basis elements of the tomography fitter, while the non-tomographic measurements used to condition the data for each fragment component will also be noisy. This means that instead of using \cref{eq:frag-tomo-probs-diag} to define our conditional probability distribution, we should use \cref{eq:frag-tomo-probs} where $\Pi_s$ is no longer diagonal, but represents our measurement error model on the conditioning qubit measurements. 

If we assume a classical readout error model these conditional qubit basis can be written as
\begin{align}
    \Pi_s^{noise} = \sum_{s^\prime} P(s|s^\prime) \ket{s^\prime}\!\!\bra{s^\prime}
\end{align}
where $P(s|s^\prime) $ is the probability of recording a true outcome $s$ as the noisy outcome $s^\prime$. The matrix of readout error probabilities $A = \sum_{s, s^\prime} P(s|s^\prime)\ket{s}\!\!\bra{s^\prime}$ is typically called an \emph{assignment matrix}~\cite{nation2021scalable}.

Using the assignment matrix readout error model our noisy conditional probabilities for fragment tomography are given by
\begin{align}
    p_{i,s}^{noise}
        = \sum_{s^\prime} P(s|s^\prime) \dbradket{B_i}{T(s^\prime)}.
\end{align}
where $B_i$ can also be chosen to be a noisy basis element corresponding to the measurement error on the cut-qubit measurements.

To apply mitigation during reconstruction we can use a modification of the conditional least-squares fitter in \cref{eq:lstsq} which simultaneously fits all fragments $T(s)$ using the readout error probabilities $P(s|s^\prime)$ using the following optimization objective:

\begin{widetext}
\begin{equation}
    \{T(s)_{LS}^{mit}\} = \mbox{arg}\min_{\{T(s)\ge 0\}} \frac12\left\|\Sigma^{-1/2}\sum_{s}\left(
    \sum_{s^\prime}
    P(s|s^\prime) \dbradket{B_i}{T(s^\prime)} - p_{i, s}\right)\ket{s}
    \right\|_2^2.
    \label{eq:lstsq-mit}
\end{equation}
\end{widetext}

\subsection{Dominant Eigenvalue Truncation (DEVT)}

Ideally a noiseless quantum circuit maps a pure state $\rho_{in}$ to another pure state $\rho_{out}$. However, in reality, the output state $\rho_{noisy}$ is a mixed state due to the effect of noise. Dominant eigenvalue truncation asserts that when the strength of the noise is low, the largest eigenvector $\ket{\psi_1}$ of $\rho_{noisy}$ has a significant overlap with $\rho_{out}$, and hence can be considered to be a very close approximation of $\rho_{out}$. The error, when the noiseless state is approximated by the largest eigenvalue of the noisy state, is captured by a quantity termed \emph{coherent mismatch}. If $\rho_{out} = \ket{\psi}\bra{\psi}$, then the coherent mismatch $c$ is defined as \cite{koczor2021dominant}
\begin{equation}
    \label{eq:coherent}
    c = 1 - \braket{\psi_1|\psi}
\end{equation}

Applicability of DEVT assumes a noise model for which the noisy output state $\rho_{noisy}$ can be represented as a convex mixture of the ideal outcome $\rho_{out}$ and some error density matrix $\rho_{err}$ as in \cref{eq:req}, $p$ being the probability of error \cite{koczor2021dominant}.
\begin{equation}
    \label{eq:req}
    \rho_{noisy} = (1-p)\rho_{out} + p\rho_{err}
\end{equation}

In such a scenario, the coherent mismatch $c$ is upper bounded as in \cref{eq:upper_bound} \cite{koczor2021dominant}, where $\delta = (\frac{1}{1-p}-1)\mu_1$, $\mu_1$ being the largest eigenvalue of $\rho_{err}$.
\begin{equation}
    \label{eq:upper_bound}
    c \leq \frac{\delta^2}{4} + \mathcal{O}(\frac{\delta^4}{16})
\end{equation}

We will focus on applying the DEVT to quantum channels as the example circuits in \cref{fig:cluster_unitary_f2q4}, \cref{fig:cluster_unitary_f2q8} and \cref{fig:cluster_unitary_3frag} contain only channel fragments. The Choi-matrix for an $n$-qubit CPTP quantum channel $\mathcal{E}$ can be written in its eigen basis as
\begin{equation}
\Lambda_{\mathcal{E}} = \sum_i \dketdbra{K_i}{K_i}
\end{equation}
where the matrices $\{K_i\}$ correspond to the canonical Kraus decomposition~\cite{wood2015qic}, and we assume they are ordered such that $\dbradket{K_i}{K_i}\ge \dbradket{K_{i+1}}{K_{i+1}}$. The DEVT approximation to $\mathcal{E}$ is then the pure-state Choi-matrix
\begin{equation}
\Lambda_{DEVT(\mathcal{E})} = \frac{2^n}{\dbradket{K_0}{K_0}} \dketdbra{K_0}{K_0}.
\end{equation}
To include this as a mitigation strategy in circuit cutting we apply the DEVT to each individual channel tensor fragment $\Lambda_{\mathcal{E}} = T_i(s_i)$ in \cref{eq:cond-tensors}, and use these truncated fragments when computing the outcome probabilities in \cref{eq:circ-probs}.

Note that when applied to quantum channels the DEVT may result in a truncated channel which is not trace-preserving, even if the non-truncated channel is. This is because a channel truncated to a single eigenvector is trace-preserving if and only if $K_0^\dagger K_0 = p\mathbb{I}$, which requires that the largest Kraus matrix be a scaled unitary $K_0 = \sqrt{p}U$.

\section{Simulations}
\label{sec:simulations}

\begin{figure}[htb]
    \centering
    \includegraphics[scale=0.3]{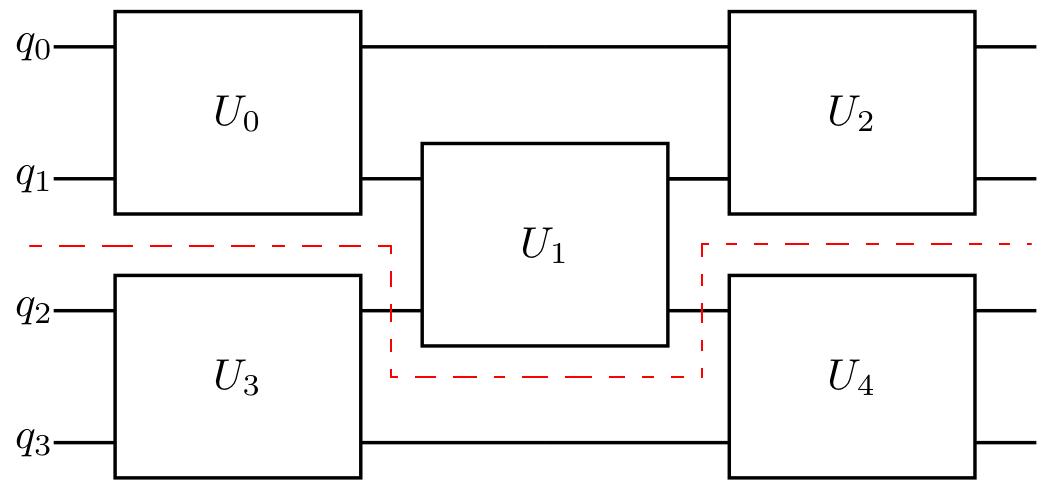}
    \caption{Cutting a 4 qubit cluster unitary circuit into 2 fragments. The dotted red line denotes the cut.}
    \label{fig:cluster_unitary_f2q4}
\end{figure}

\begin{figure}[htb]
    \centering
    \includegraphics[scale=0.25]{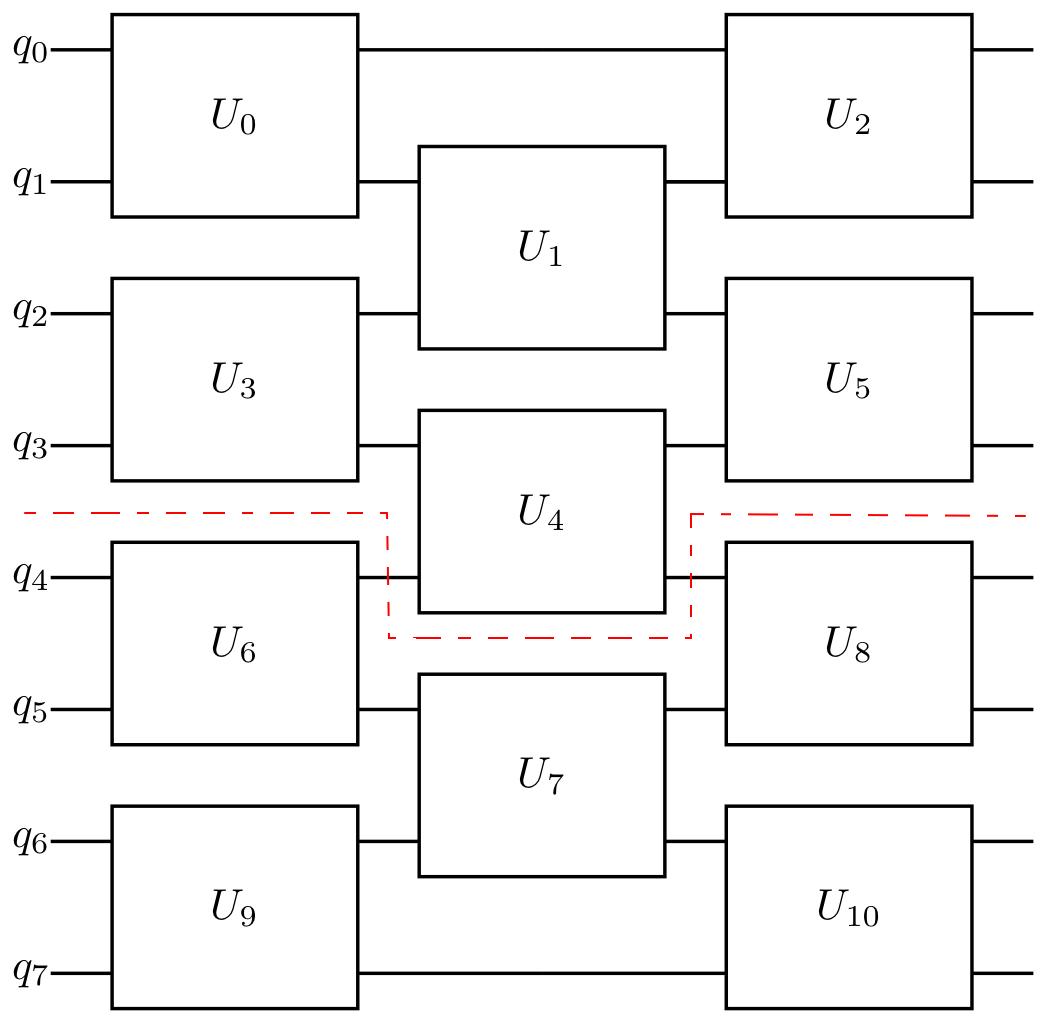}
    \caption{Cutting a 8 qubit cluster unitary circuit into 2 fragments. The dotted red line denotes the cut.}
    \label{fig:cluster_unitary_f2q8}
\end{figure}

\begin{figure}[htb!]
    \centering
    \includegraphics[scale=0.3]{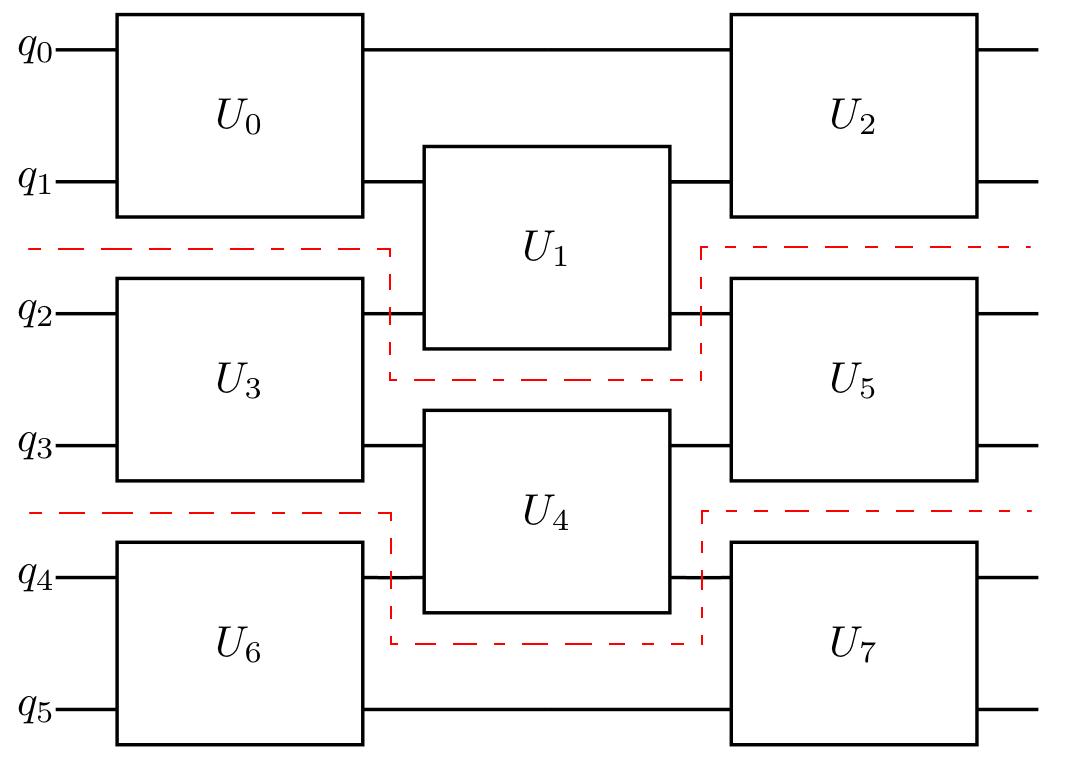}
    \caption{Cutting a 6 qubit cluster unitary circuit into 3 fragments. The dotted red line denotes the cut.}
    \label{fig:cluster_unitary_3frag}
\end{figure}

For our numerical study we consider a cluster unitary circuit consisting of alternating layers of random 2-qubit unitary gates. Such circuits are representative of Trotterized simulation used in applications for near-term quantum devices \cite{berg2022probabilistic, cerezo2021variational}. In our experiments we consider clusters with a fixed depth of 3 layers of random unitaries between adjacent qubits and the problem of estimating the full outcome probability distribution of Z-basis measurements on all qubits. We compare direct simulation of the full uncut circuit to the simplest circuit cutting configuration with 2 or 3 fragments, as shown in \cref{fig:cluster_unitary_f2q4,fig:cluster_unitary_f2q8,fig:cluster_unitary_3frag}, under a variety of noise models. For 2-fragment comparison simulations were performed for original uncut circuits containing 4 qubits (\cref{fig:cluster_unitary_f2q4}), 8 qubits (\cref{fig:cluster_unitary_f2q8}) and 12 qubits, resulting in 1-qubit conditional process tomography fragment tensors with 2, 4, and 6 conditioning qubit measurements from the uncut circuit respectively. In \cref{fig:cluster_unitary_3frag} we also show an example of cutting a 6 qubit cluster unitary into 3 fragments.

To evaluate the performance of circuit cutting compared to direct simulation we compare the estimated probability distribution from both methods to the expected ideal distribution using the 1-norm distance measure
\begin{align}
D(P) =  \frac{1}{2} \|P - Q\|_1 =  \frac{1}{2} \sum_i |p_i - q_i|
\label{eq:trace-distance}
\end{align}
where $p_i$ and $q_i$ are the outcome probabilities for the estimated distribution $P$, and ideal target distribution $Q$ respectively. Note that this measure is equivalent to the \emph{trace distance} $T(\rho, \sigma) = \frac{1}{2}\mbox{Tr}|\rho - \sigma|$ between two quantum states if we consider the probability distributions to be diagonal density matrices. 

For performing the circuit cutting reconstruction we compared 3 different tomography fitters as described in \cref{sec:frag-tomo}: linear inversion (LIN), constrained least-squares (CLS), and measurement error mitigated constrained least-squares (MEMCLS). For each fitter we compared with and without dominant Eigenvalue truncation (DEVT) error mitigation, and also considered the effectiveness of partial tomography where the fitter was run using only a subset of collected measurement data. All tomography reconstruction experiments consisted of $10000$ trials, or \emph{shots}, and were implemented using a modified version of the process-tomography experiment from the \emph{Qiskit Experiments} \cite{qexp} Python package. For comparison with direct simulation, we also included the effect of standard A-matrix inversion readout error mitigation on the estimated probabilities using the \emph{M3} mitigation package in Qiskit~\cite{nation2021scalable,mthree}.

To study the effects or error mitigation we compared a variety of noise models. All noisy simulations were done using the \emph{Qiskit Aer} simulator \cite{aer} with a local noise model by decomposing the simulated circuits into Controlled-NOT, and 1-qubit gates (SX, X, RZ, which is currently the basis gate set of IBM quantum devices), and Z-basis measurements. Noise was then added to either the 2-qubit gates, 1-qubit gates, or single-qubit measurements using the same noise parameters for all qubits to simplify analysis. Note that in IBM Quantum devices, RZ gate is not physically executed, rather its effect is accounted for in the software by a rotation of axis \cite{PhysRevA.96.022330}. Hence, RZ is essentially a virtual gate causing no error. Therefore, single qubit gate error was added to X and SX gates only. 

Measurement errors were simulated using the classical readout error model as described in \cref{sec:meas-noise}, where we assumed a \emph{local} noise model where the readout error of each qubit is uncorrelated with other qubits. For gate errors we use a local Markovian gate model where each noisy gate is simulated as $\mathcal{U}_{noise} = \mathcal{E} \cdot U$ where $U$ is the ideal gate unitary, and $\mathcal{E}$ is a completely-positive trace preserving (CPTP) quantum noise channel, which can be written in the Kraus representation as $\mathcal{E}(\rho)=\sum_i K_i \rho K_i^\dagger$, where $\sum_i K_i^\dagger K_i = \mathbb{I}$.~\cite{wood2015qic}. We simulated four different gate noise models for the channel $\mathcal{E}$ (i) depolarizing, (ii) Pauli, (iii) amplitude damping, and (iv) coherent noise, where noise was applied to both 1 and 2 qubit gates. While we only consider local gate noise for simplicity of the analysis, we note that more realistic noise such as non-local cross-talk fall into these categories as non-local coherent errors, or Pauli error channels if techniques such as Pauli-twirling are applied to convert coherent noise to incoherent Pauli noise~\cite{wallman2016rc}.

\begin{figure*}[htb]
    \centering
    \subfigure[2 fragment circuit cutting with $p_{\scriptsize{meas}}=0.01$]{
        \includegraphics[scale=0.35]{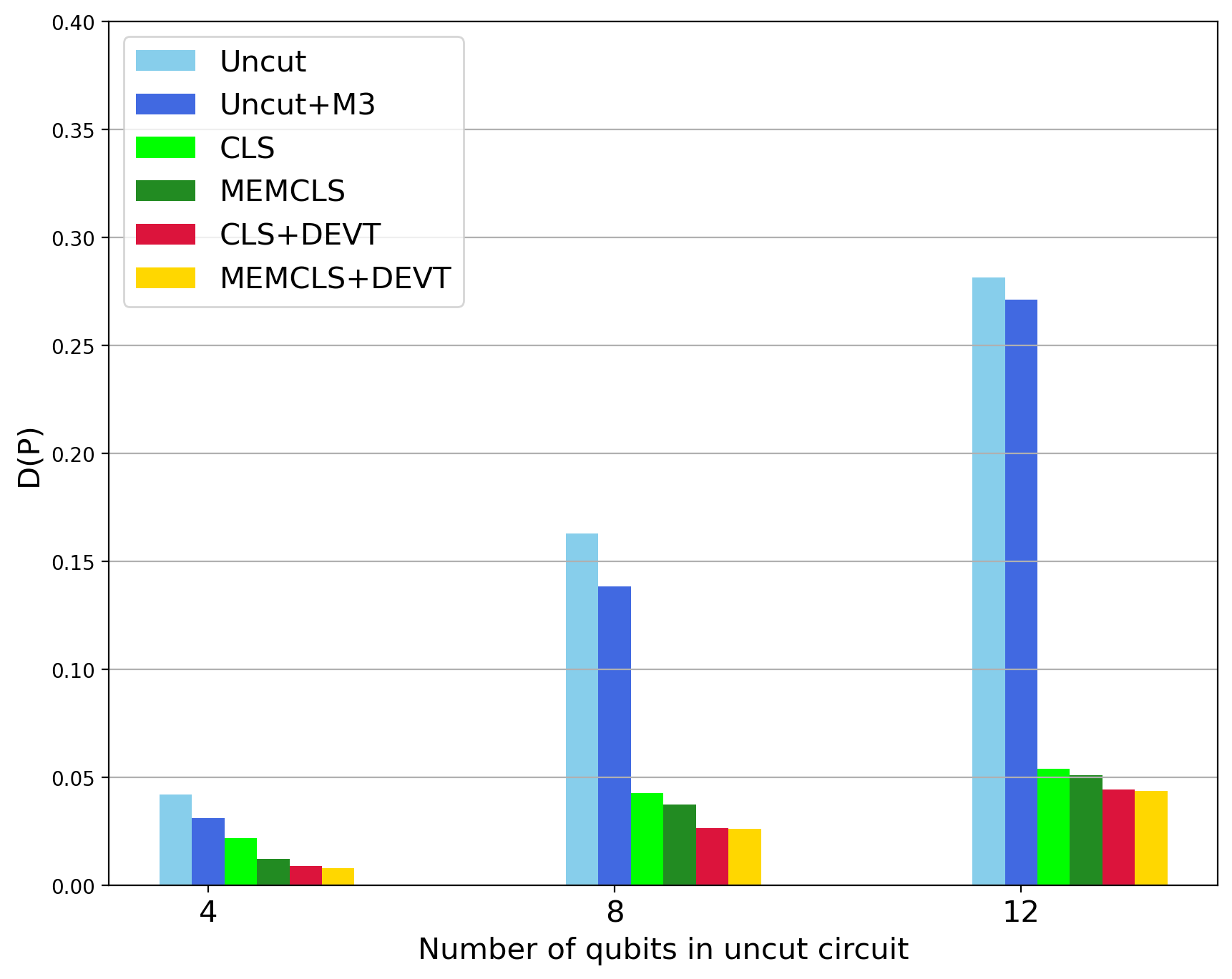}
    }\label{fig:mem-2f_01}
    \subfigure[2 fragment circuit cutting with $p_{\scriptsize{meas}}=0.05$]{
        \includegraphics[scale=0.35]{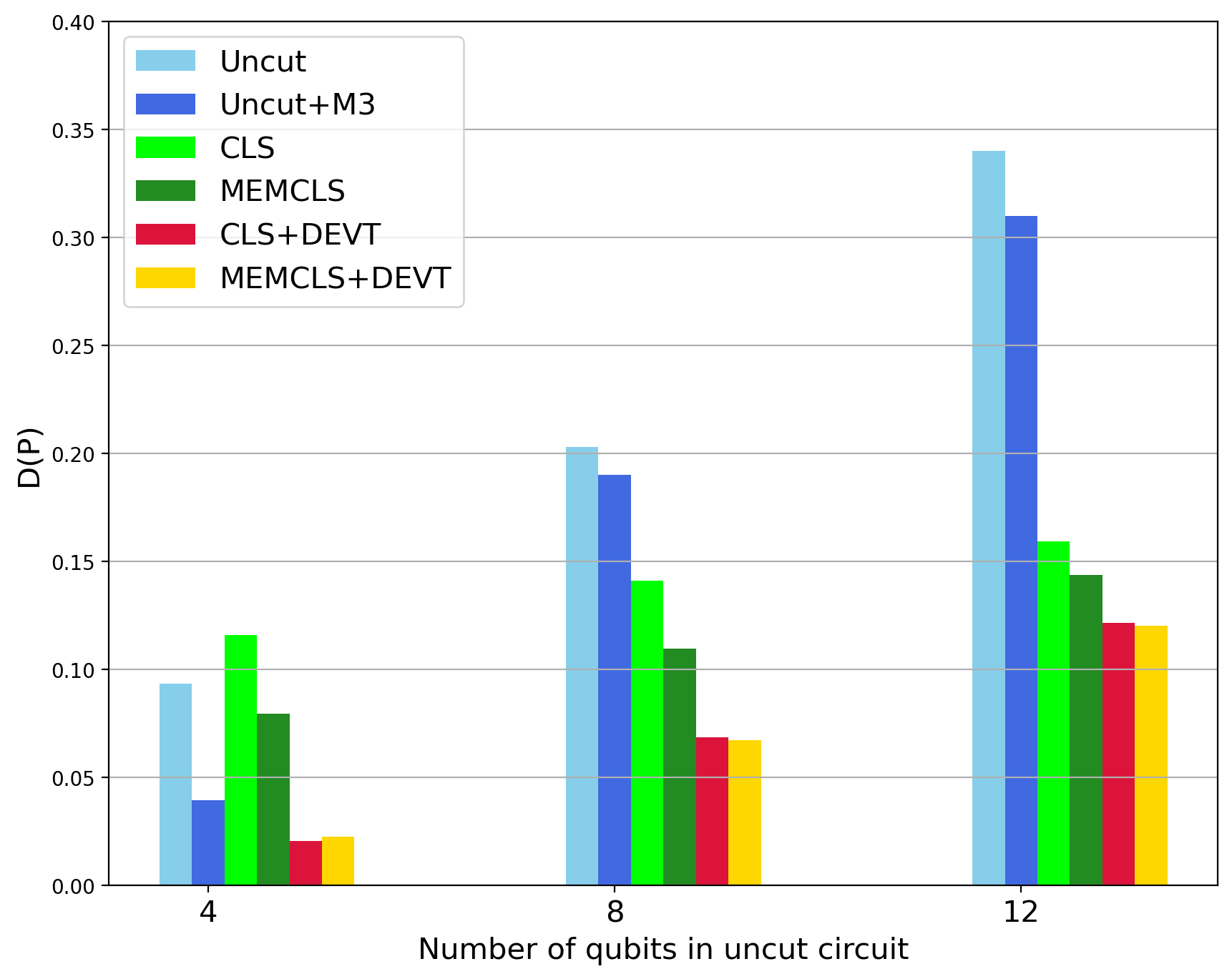}
    }\label{fig:mem-2f_05}
    \subfigure[3 fragment circuit cutting with $p_{\scriptsize{meas}}=0.01$]{
        \includegraphics[scale=0.35]{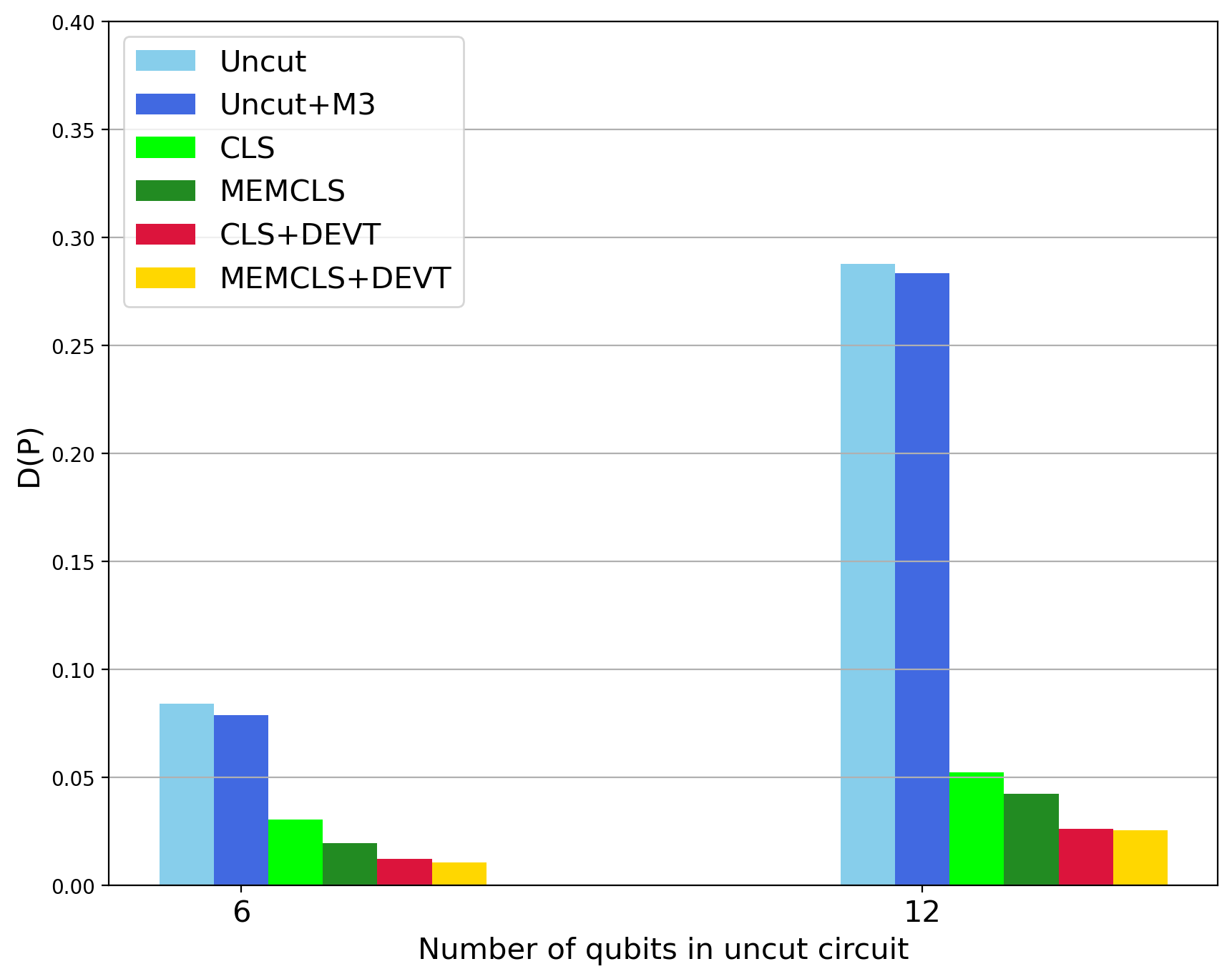}
    }\label{fig:mem-3f_01}
    \subfigure[3 fragment circuit cutting with $p_{\scriptsize{meas}}=0.05$]{
        \includegraphics[scale=0.35]{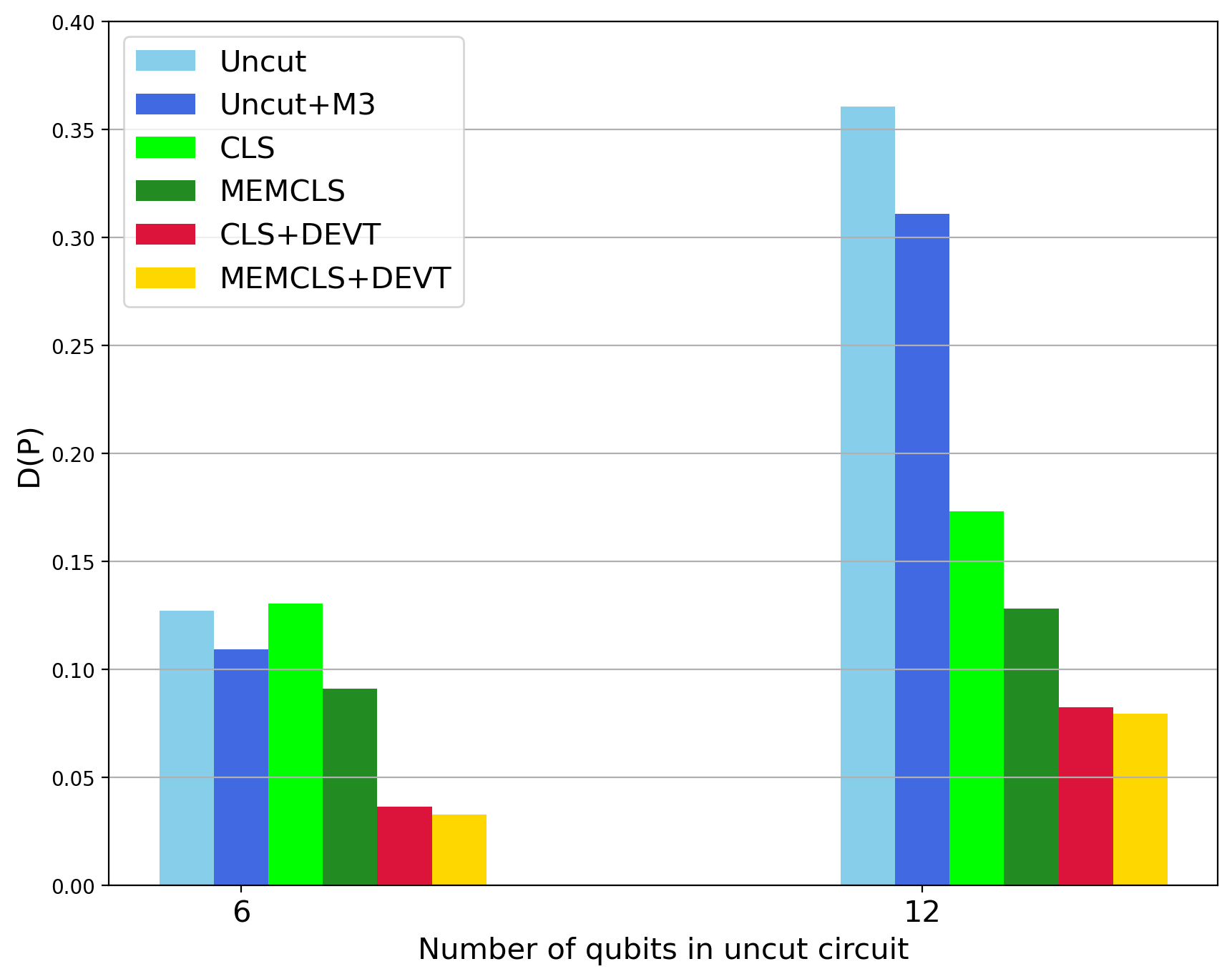}
    }\label{fig:mem-3f_05}
    \caption{Performance of tomographic circuit cutting reconstruction using 2 and 3 fragments under the effect of local symmetric readout error with a readout error probability of $p_{\scriptsize{meas}} \in \{0.01, 0.05\}$. The vertical axis is the trace distance (\cref{eq:trace-distance}) of the reconstructed probability distribution from the noiseless probability distribution. For the cut circuit we performed reconstruction using both constrained least-squares conditional tomography fitter (CLS) and a readout error error mitigated fitting (MEMCLS) fitter using the noisy basis corresponding to the classical readout error noise parameter both with and without DEVT mitigation. The original circuit (uncut) was measured with and without M3 readout error mitigation for comparsion.
    }
    \label{fig:mem}
\end{figure*}

\begin{figure*}
    \centering
    \subfigure[$p_{\scriptsize{depol}} = 0.01$, $p_{\scriptsize{meas}} = 0.01$]{
        \includegraphics[scale=0.35]{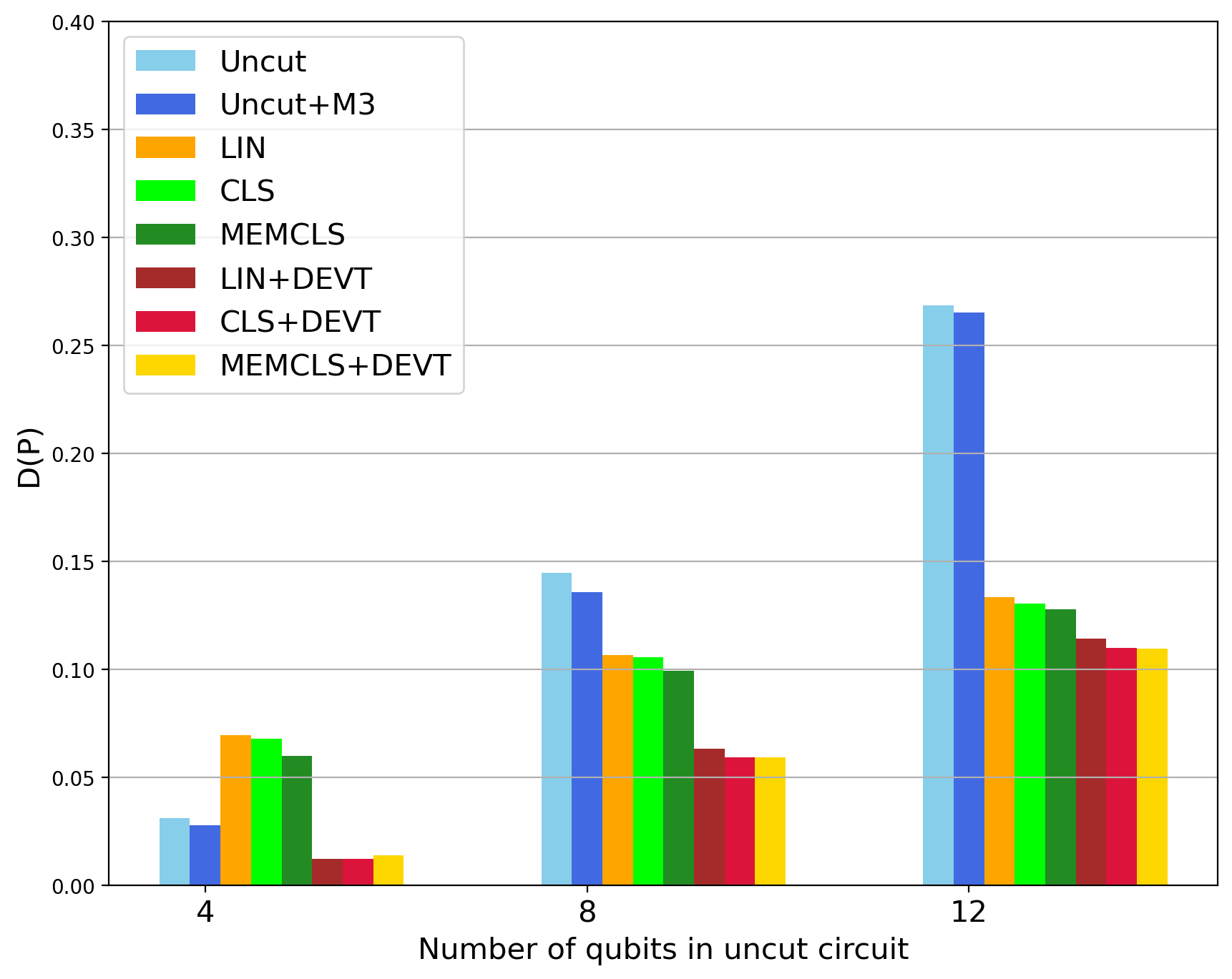}
    }\label{fig:depol-a}
    \subfigure[$p_{\scriptsize{depol}} = 0.02$, $p_{\scriptsize{meas}} = 0.01$]{
        \includegraphics[scale=0.35]{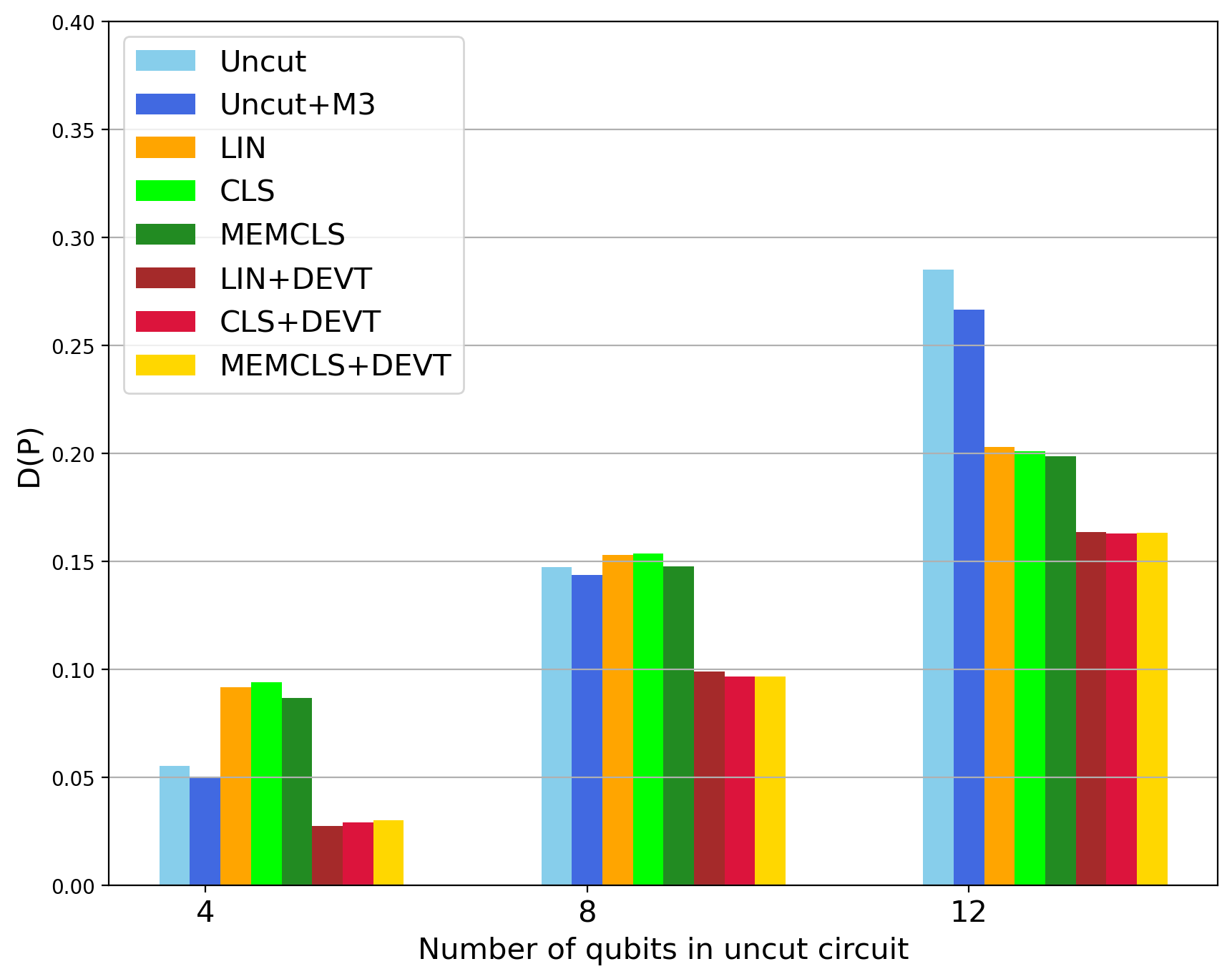}
    }\label{fig:depol-b}
    \subfigure[$p_{\scriptsize{depol}} = 0.01$, $p_{\scriptsize{meas}} = 0.05$]{
        \includegraphics[scale=0.35]{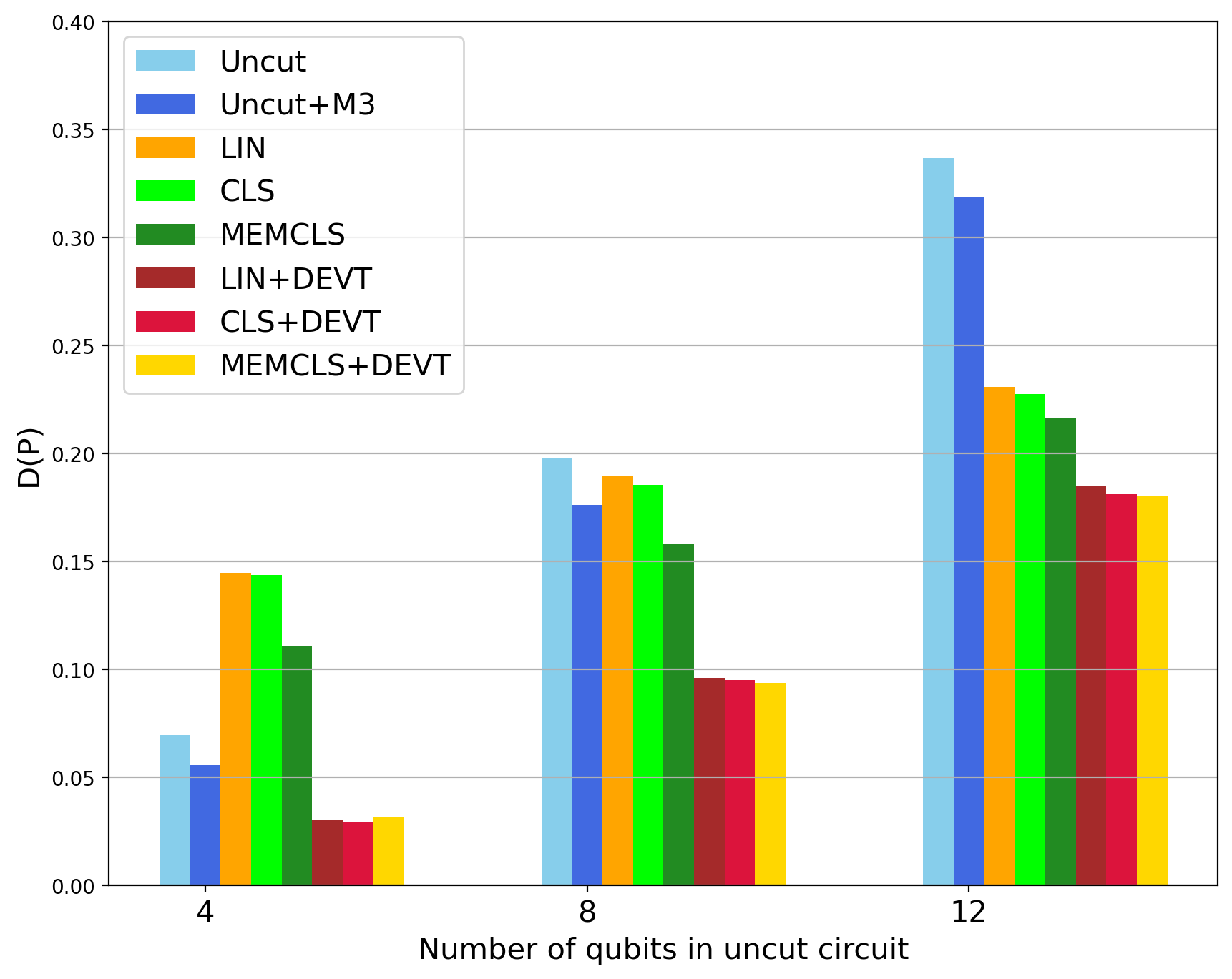}
    }\label{fig:depol-c}
    \subfigure[$p_{\scriptsize{depol}} = 0.02$, $p_{\scriptsize{meas}} = 0.05$]{
        \includegraphics[scale=0.35]{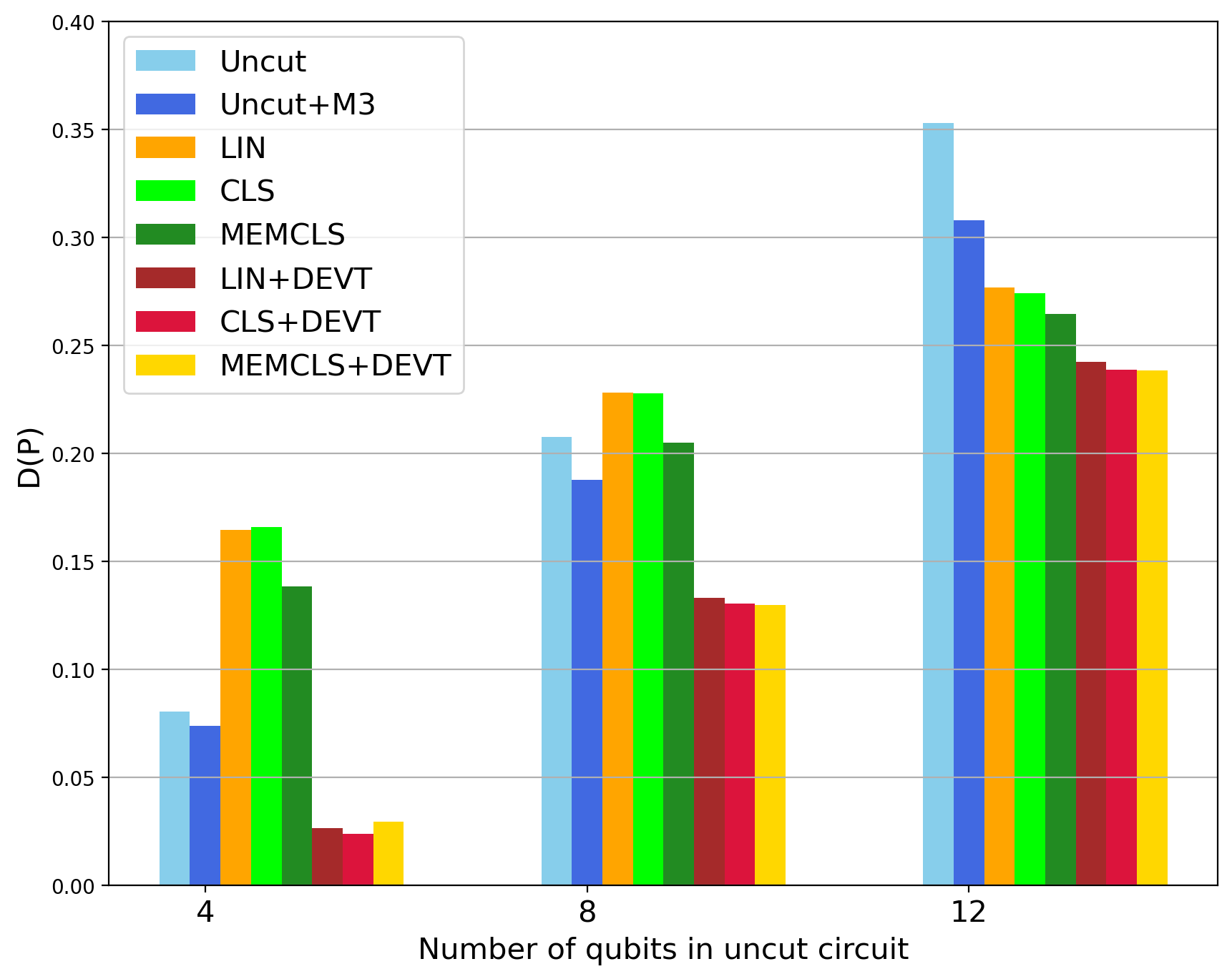}
    }\label{fig:depol-d}
    \caption{Performance of error mitigated 2-fragment tomographic circuit cutting reconstruction under 2-qubit depolarizing gate noise (\cref{eq:depol_channel}) with $p_{\scriptsize{depol}}=0.01$ (left) or $p_{\scriptsize{depol}} = 0.02$ (right), local symmetric readout error with $p_{\scriptsize{meas}} = 0.01$ (top) or $p_{\scriptsize{meas}} = 0.05$ (bottom), and single qubit gate depolarizing error of $p_1 = 10^{-4}$. The two-qubit depolarizing noise parameter was Errors on single qubit gates are fixed to $10^{-4}$. Cut circuit reconstruction was compared using linear inversion (LIN), constrained least-squares (CLS), and readout error mitigated CLS (MEMCLS) tomography fitters both with and without dominant eigenvalue truncation (DEVT) mitigation. The original circuit (uncut) was measured with and without M3 readout error mitigation for comparison.}
    \label{fig:depol}
\end{figure*}

\subsection{Measurement Noise}
\label{sec:meas-noise}

First we simulated using measurement noise with ideal gates and a symmetric single-qubit assignment matrix
\begin{equation}
A = \begin{pmatrix}
    1 - p_{\scriptsize{meas}} & p_{\scriptsize{meas}}  \\
    p_{\scriptsize{meas}} & 1 - p_{\scriptsize{meas}}
\end{pmatrix}.
\end{equation}
\cref{fig:mem} shows the reconstructed distributions trace distance for CLS, MEMCLS both with and without DEVT, and the uncut circuit both with and without M3 readout error mitigation when using $p_{\scriptsize{meas}} \in \{0.01, 0.05\}$. For $p_{\scriptsize{meas}} = 0.01$, we find that circuit cutting alone improves the performance for all the scenarios. When $p_{\scriptsize{meas}}=0.05$, for 8-qubit and 12-qubit 2-fragment circuits, we find that circuit cutting alone improves performance, as was reported in prior studies \cite{ayral2021quantum, basu2021qer}. This improvement increases with the number of measurement qubits in the original circuit, while for the 4-qubit case the opposite is seen with the uncut circuit having lower error than the cut circuit reconstruction. This is because for the 4-qubit case the total number of outcome probabilities used in each fragment for the tomographic reconstruction is larger than the number of probabilities in the original circuit (24 vs 16), while for the 8 and 12-qubit cases each fragment has significantly fewer probabilities than the original circuit (96 vs 256 and 384 vs 4096 respectively).

We find that including readout error mitigation using MEMCLS or CLS with DEVT improves the circuit cutting performance, with CLS+DEVT outperforming MEMCLS alone, and being comparable to MEMCLS+DEVT. This suggests DEVT is more effective than MEMCLS for mitigating pure readout errors in tomographic reconstructions which we discuss more in \cref{app:devt-meas}. The same trends are also observed for reconstruction using 3 fragments shown in \cref{fig:mem} (c) and (d). Henceforth, for other noise models, we show the results for circuit cutting with 2 fragments only, since for equal sized fragments the reconstruction error per fragment is constant and the 2-fragment case is representative of relative performance of the different methods.

\begin{figure*}
    \centering
    \subfigure[Two qubit gate error $p = 0.01$ and bias $b = 0.1$]{
        \includegraphics[scale=0.35]{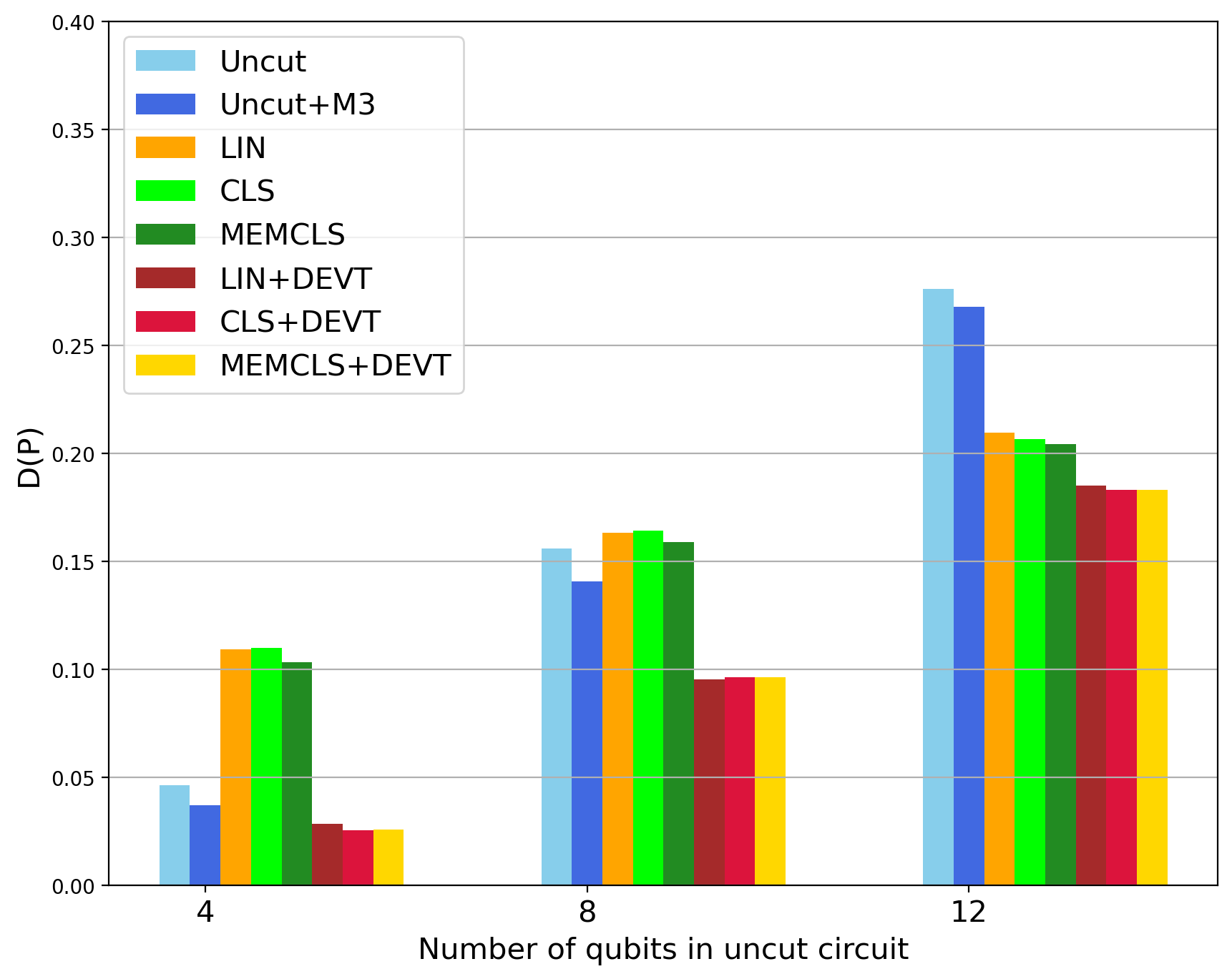}
    }\label{fig:pauli-a}
    \subfigure[Two qubit gate error $p = 0.02$ and bias $b = 0.1$]{
        \includegraphics[scale=0.35]{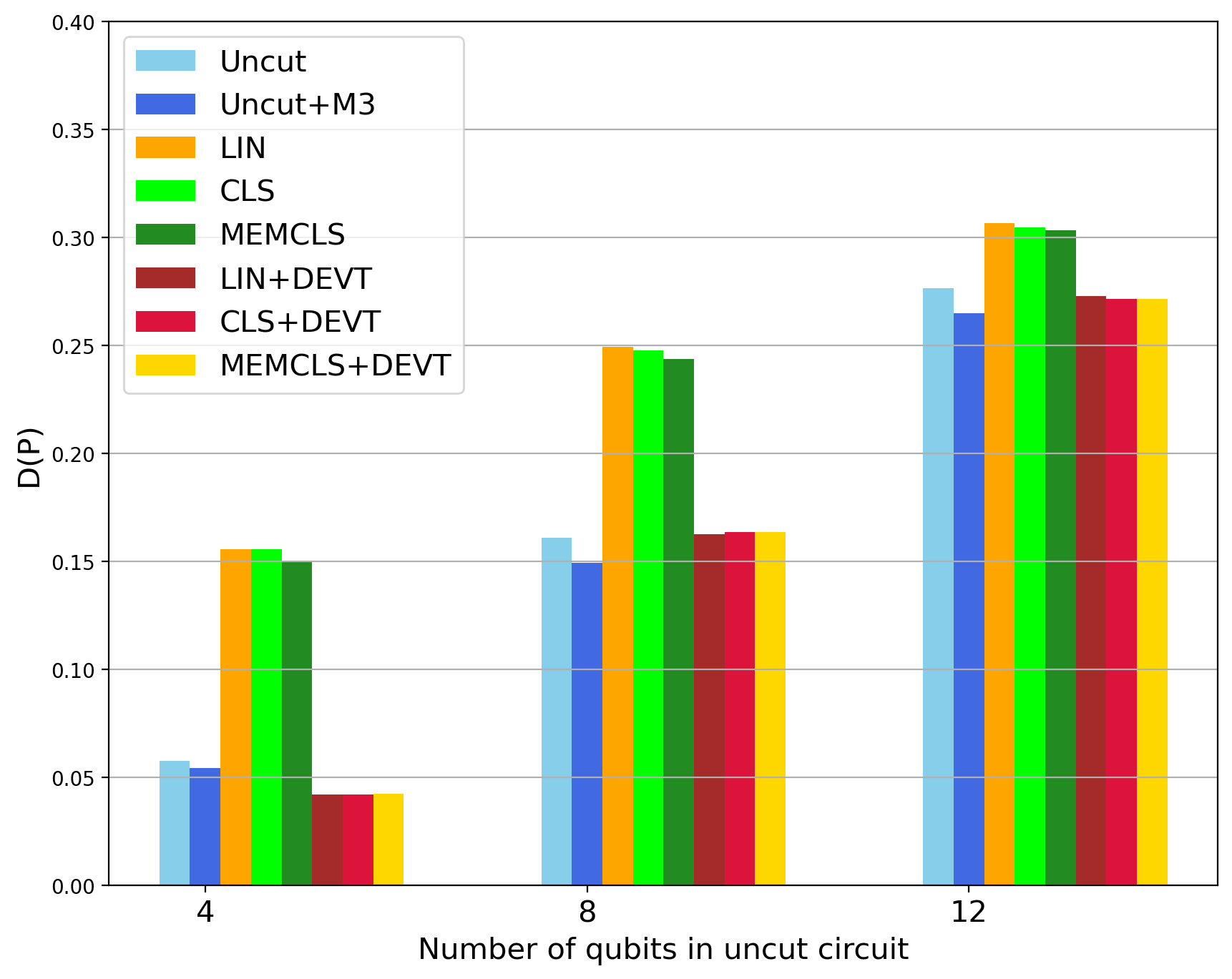}
    }\label{fig:pauli-b}
    \subfigure[Two qubit gate error $p = 0.01$ and bias $b = 0.5$]{
        \includegraphics[scale=0.35]{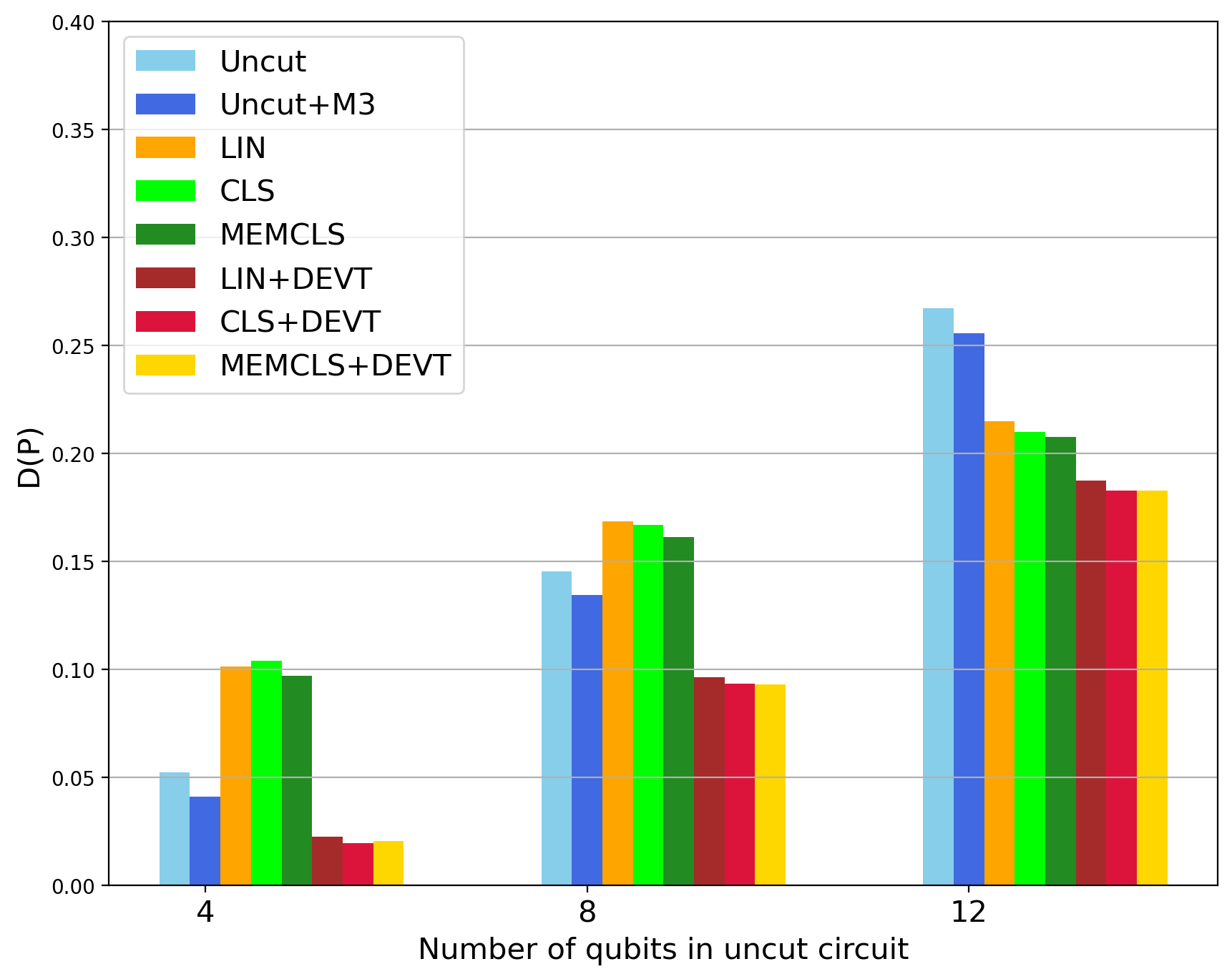}
    }\label{fig:pauli-c}
    \subfigure[Two qubit gate error $p = 0.02$ and bias $b = 0.5$]{
        \includegraphics[scale=0.35]{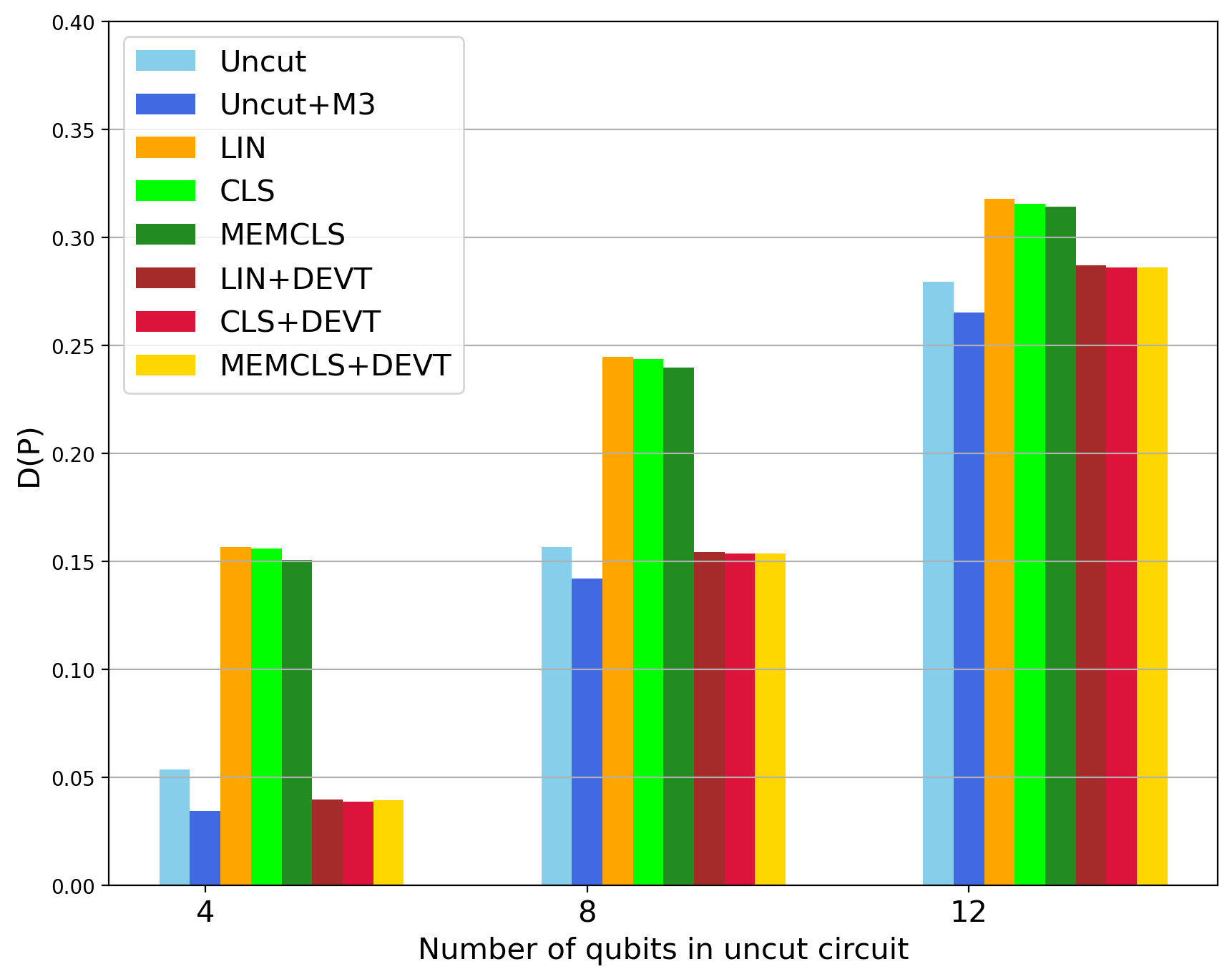}
    }\label{fig:pauli-d}
    \caption{Performance of error mitigated 2-fragment tomographic circuit cutting reconstruction under a 2-qubit tensor product biased Pauli error channel (\cref{eq:pauli_noise}) with $p_X = p_Y = p$, and $p_z = p(1 + b)$ with probabilities $p=0.01$ (left) and $p=0.02$ (right), and biases $b=0.1$ (top) and $b=0.5$ (bottom), local symmetric readout error with $p_{\scriptsize{meas}} = 0.05$, and single qubit gate depolarizing error of $p_1 = 10^{-4}$. Cut circuit reconstruction was compared using linear inversion (LIN), constrained least-squares (CLS) tomography fitters, readout error mitigated CLS (MEMCLS) both with and without dominant eigenvalue truncation (DEVT) mitigation. The original circuit (uncut) was measured with and without M3 readout error mitigation for comparison.}
    \label{fig:pauli}
\end{figure*}

\subsection{Gate Noise}
\label{sec:gate-results}

Next we include the effects of gate error as well as measurement readout error. In current devices 2-qubit gate error is one of the dominant sources of circuit error, so we consider several different 2-qubit noise models applied to all CNOT gates in our circuit. In all cases we fix the measurement error to be the symmetric classical readout error described in \cref{sec:meas-noise} with $p_{\scriptsize{meas}}=0.01$ or $0.05$, and including a 1-qubit gate error model (depolarizing or pauli) with $p_1 = 10^{-4}$ on single-qubit gates.

The first case we consider is a 2-qubit depolarizing noise channel given by the map
\begin{equation}
    \mathcal{E}_{\scriptsize{depol}}(\rho) = (1-p_{\scriptsize{depol}}) \rho + p_{\scriptsize{depol}} \frac{\mathbb{I}}{4}.
    \label{eq:depol_channel}
\end{equation}

We simulated with 2-qubit depolarizing probabilities of $p_{\scriptsize{depol}}=0.01$ and $0.02$, shown respectively in \cref{fig:depol}. We find that both CLS and LIN tomography fitters with DEVT perform better than MEMCLS alone, and that when applying DEVT mitigation all tomography fitters have comparable performance. This suggests that if one is employing DEVT, then full MEMCLS is not required over standard CLS or basic LIN tomography fitting, since DEVT alone is sufficient to mitigate the effect of both depolarizing gate error and measurement readout error on quantum circuits. However, as the size of the fragments increase, so does the number of conditional qubits and the effect of measurement error on them, making measurement error more and more dominant. Thus the performance gap between MEMCLS and CLS with DEVT lowers with increasing size of fragments. Moreover, numerical data suggests that the rate of increase of $\text{trace distance}$ increases with the increase in $p_{\scriptsize{depol}}$ and the number of qubits in the fragment, which is expected. If one considers that circuit error in a layer can be approximated as a depolarizing error, then $1-D(P)$ obtained using DEVT would scale as $\mathcal{O}(nmp)^2$, where $n$ is the number of qubits, $m$ is the number of gate layers, and $p$ is the layer depolarizing error probability (See \cref{app:devt-depol} for details).

Depolarizing noise was expected to be the most favourable case for DEVT. However it is not a realistic model for most quantum devices. A more general mixed-unitary model is a general Pauli channel with different error rates. We consider a biased Pauli channel where the Z error term is more likely than the X or Y terms. This is given by the map
\begin{equation}
    \label{eq:pauli_noise}
    \mathcal{E}_{\mbox{\scriptsize{pauli}}}(\rho) = (1-(3+b)p)\rho + p X\rho X + p Y \rho Y + p(1+b) Z \rho Z
\end{equation}
where $p$ is the X and Y error probability, and $b$ is a bias term added to the $Z$ error. We simulated with values of $p=0.01$ and $0.02$ in \cref{fig:pauli} (a) and (b) respectively with a bias term of $b = 0.1$, and in \cref{fig:pauli} (c) and (d) respectively with a bias term of $b = 0.5$. An $n$-qubit Pauli noise channel is defined as the tensor product of individual qubit noise channel $(\mathcal{E}_{\mbox{\scriptsize{pauli}}})^{\otimes n}$.

For 2-qubit gates with an error probability of $0.01$, for both tomography fitters and both values of bias, we find that while DEVT does not improve the results as much as for depolarizing noise, it still provides an advantage over both the uncut circuit, and the cut circuit without DEVT. However, when the probability of gate error is increased to $0.02$, for a bias of $0.1$, we notice that DEVT hardly provides any improvement over the uncut circuit. In fact, for the 12 qubit circuit, when the bias is $0.5$, we find that even DEVT is providing a result which is slightly worse than that of the uncut circuit. We show, in \cref{app:devt-pauli}, that the form of the output density matrix for biased Pauli noise model deviates from \cref{eq:req}, leading to a poorer result as compared to uniform depolarization noise model. There we also show analytically that DEVT on biased quantum channels are unlikely to provide a performance as good as that of the depolarizing channel. However, these numerical results clearly suggest that DEVT provides an improvement over circuit cutting without DEVT in every scenario, thus consolidating its necessity for circuit cutting with noise.

In both \cref{fig:depol} and \cref{fig:pauli} we observe that the reconstruction error increases by a larger amount when increasing the gate noise parameter for all tomographic circuit cutting methods than is observed when measuring the uncut circuit. This effect was consistent across all noise models considered in following sections.

\begin{figure*}
    \centering
    \subfigure[Amplitude damping error with $\gamma = 0.001$]{
        \includegraphics[scale=0.35]{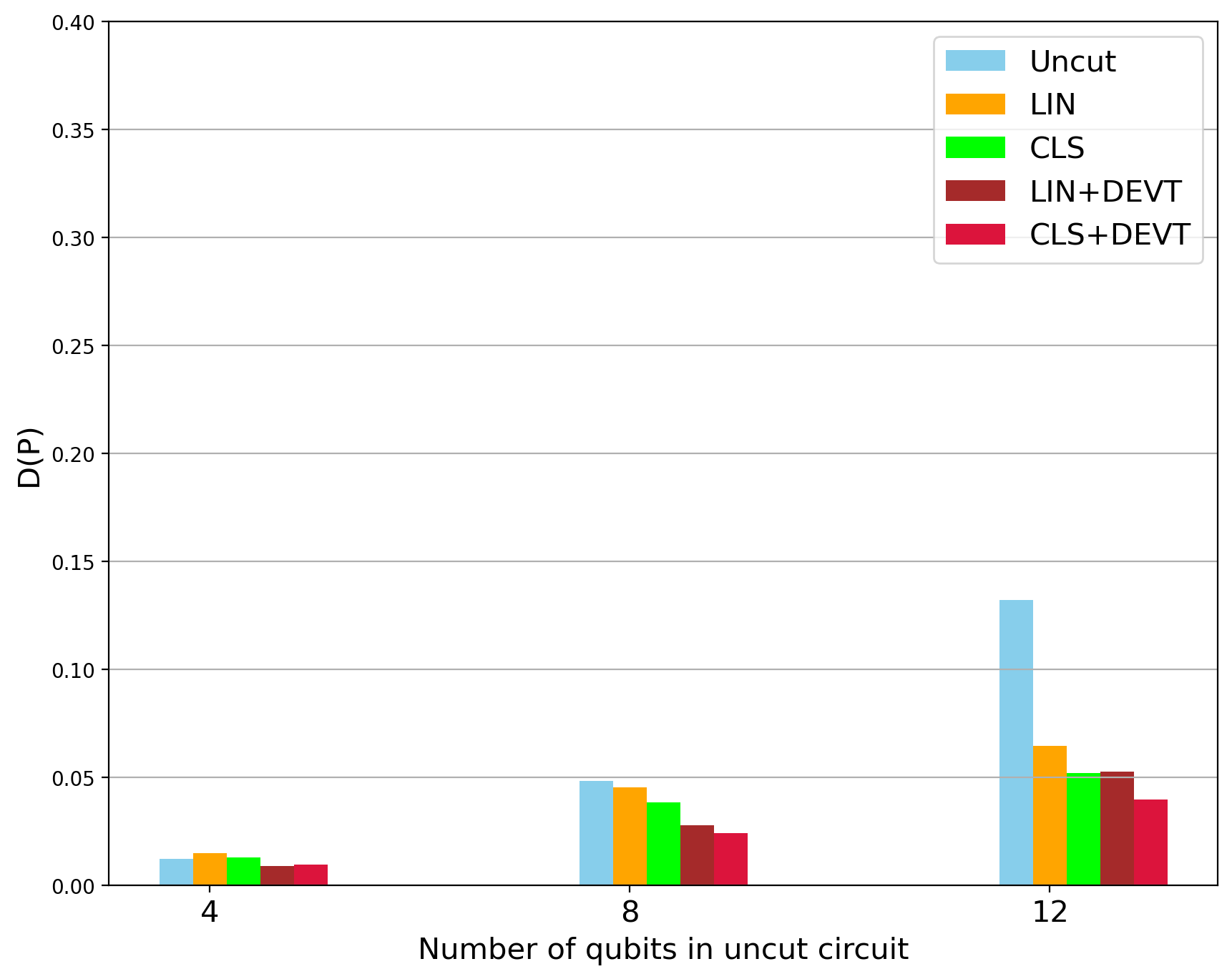}
    }\label{fig:ad_a}
    \subfigure[Amplitude damping error with $\gamma = 0.01$]{
        \includegraphics[scale=0.35]{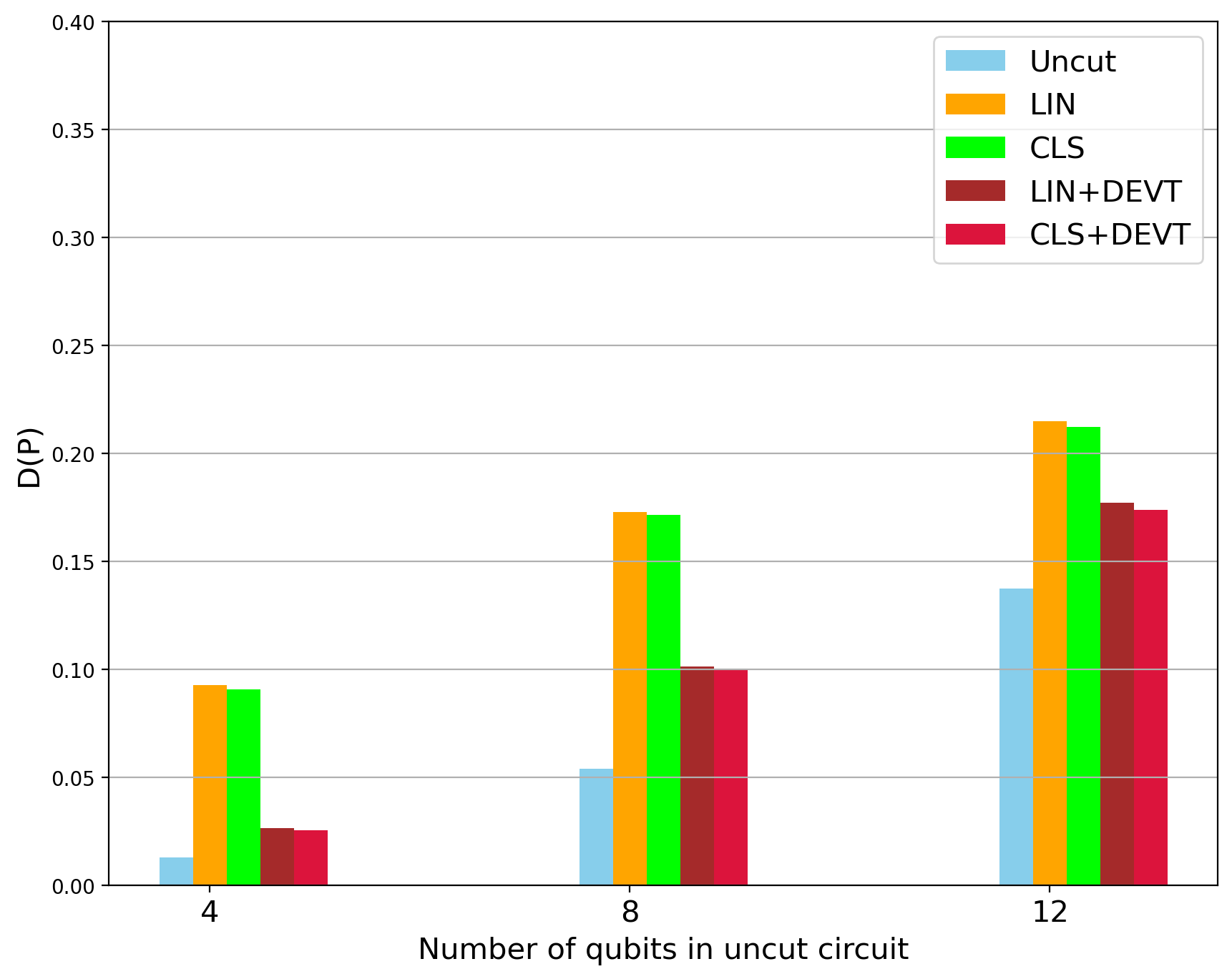}
    }\label{fig:ad_b}
    \caption{Performance of error mitigated 2-fragment tomographic circuit cutting reconstruction under a 2-qubit tensor product amplitude damping error channel (\cref{eq:ad_channel}) with damping parameter $\gamma=0.001$ (a), and $\gamma=0.01$ (b). Cut circuit reconstruction was compared using linear inversion (LIN) and constrained least-squares (CLS) tomography fitters, both with and without dominant eigenvalue truncation (DEVT) mitigation. Direct measurement of the original circuit (uncut) is shown for comparison.}
    \label{fig:ad}
\end{figure*}

\begin{figure*}
    \centering
    \subfigure[Coherent error with $\Delta \theta = \frac{\pi}{64}$]{
        \includegraphics[scale=0.35]{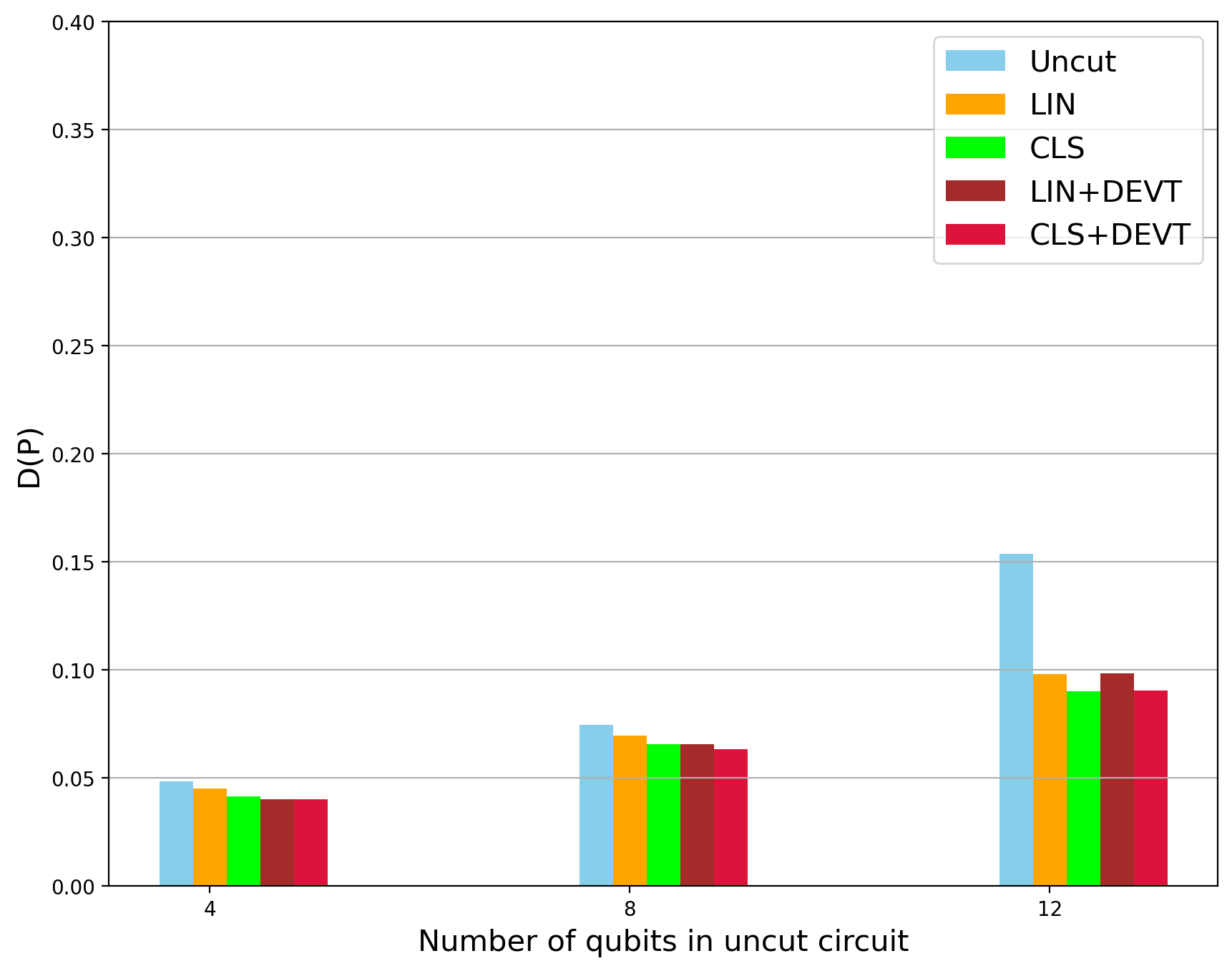}
    }\label{fig:coherent_a}
    \subfigure[Coherent error with $\Delta \theta = \frac{\pi}{32}$]{
        \includegraphics[scale=0.35]{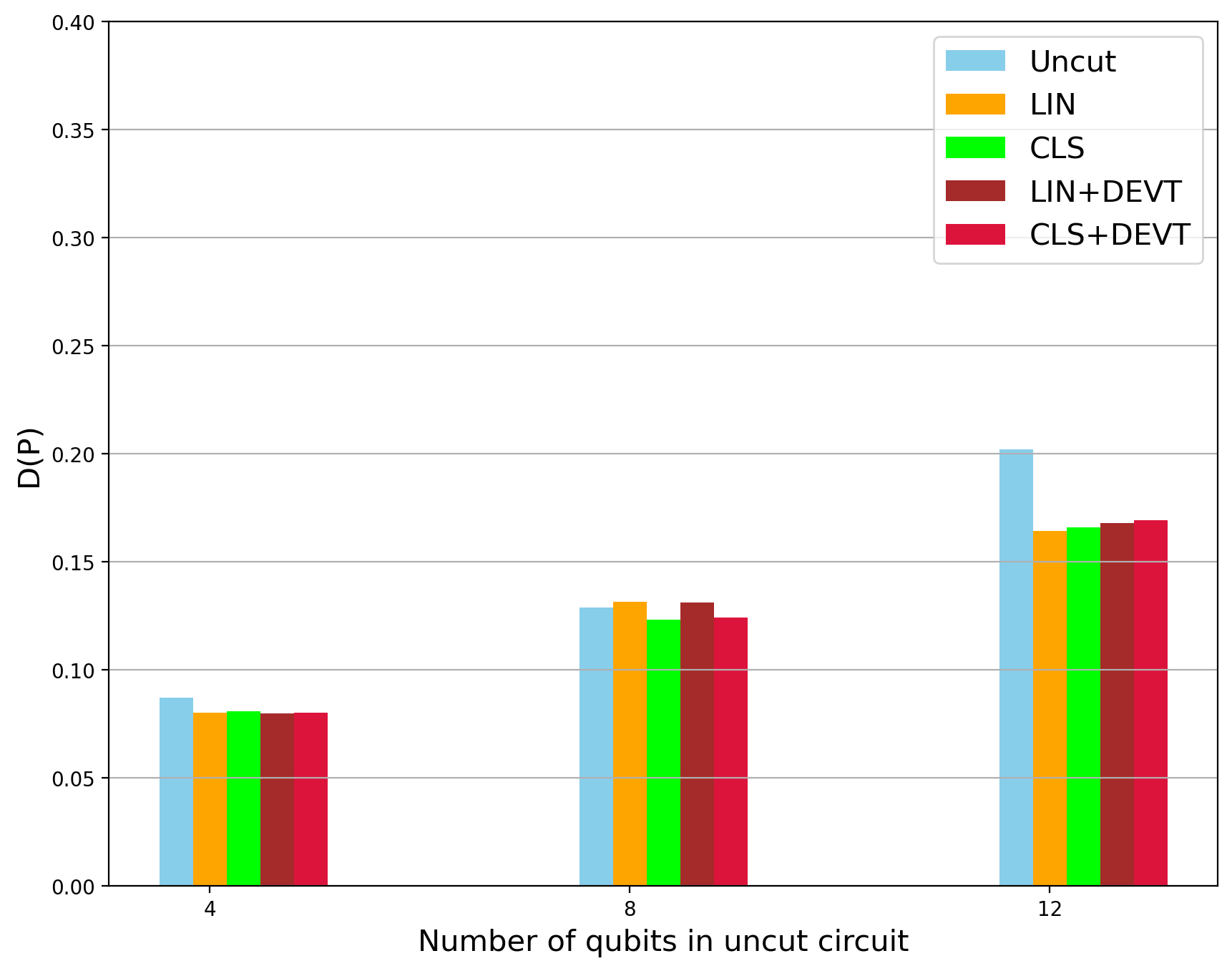}
    }\label{fig:coherent_b}
    \caption{Performance of error mitigated 2-fragment tomographic circuit cutting reconstruction under a 2-qubit tensor product coherent rotation error channel (\cref{eq:coh_error}) with rotation error $\Delta\theta=\pi/64$ (a), and $\Delta\theta=\pi/32$ (b). Cut circuit reconstruction was compared using linear inversion (LIN) and constrained least-squares (CLS) tomography fitters, both with and without dominant eigenvalue truncation (DEVT) mitigation. Direct measurement of the original circuit (uncut) is shown for comparison.}
    \label{fig:coherent}
\end{figure*}

\subsection{Non-Mixed Unitary Gate Errors}\label{sec:non-pauli}
Next we consider two other representative cases of non-mixed-unitary error, amplitude damping and coherent noise, both of which result in an error map not of the form in \cref{eq:req}, and hence should be unfavourable for DEVT. For these simulations we do not include measurement readout error so as to asses DEVT for these gate errors without including improvement from its effectiveness for mitigating measurement readout errors.

First we consider 2-qubit gate error consisting of a tensor product of 1-qubit amplitude damping channels $\mathcal{E}_{\scriptsize{amp}}(\rho) = K_0\rho K_0^\dagger + K_1 \rho K_1^\dagger$ with
\begin{equation}
    \label{eq:ad_channel}
    K_0 = \begin{pmatrix}
1 & 0 \\
0 & \sqrt{1- \gamma}
\end{pmatrix} \quad K_1 = \begin{pmatrix}
0 & \sqrt{\gamma} \\
0 & 0
\end{pmatrix}
\end{equation}
where we consider damping parameter values of $\gamma=0.001$, shown in \cref{fig:ad} (a), and $\gamma=0.01$, shown in \cref{fig:ad} (b). We observe that for $\gamma=0.001$, when the channel can be considered to be very close to identity, DEVT, and circuit cutting in general, are able to attain improved performance. For $\gamma=0.01$, both LIN and CLS with DEVT provides an improvement over either fitter without DEVT, however, it is not able to match the performance of the full circuit, and in fact the cut circuit reconstruction appears more sensitive to the gate errors than the uncut circuit.

\begin{figure*}
    \centering
    \subfigure[Amplitude damping error with $\gamma = 0.01$]{
        \includegraphics[scale=0.35]{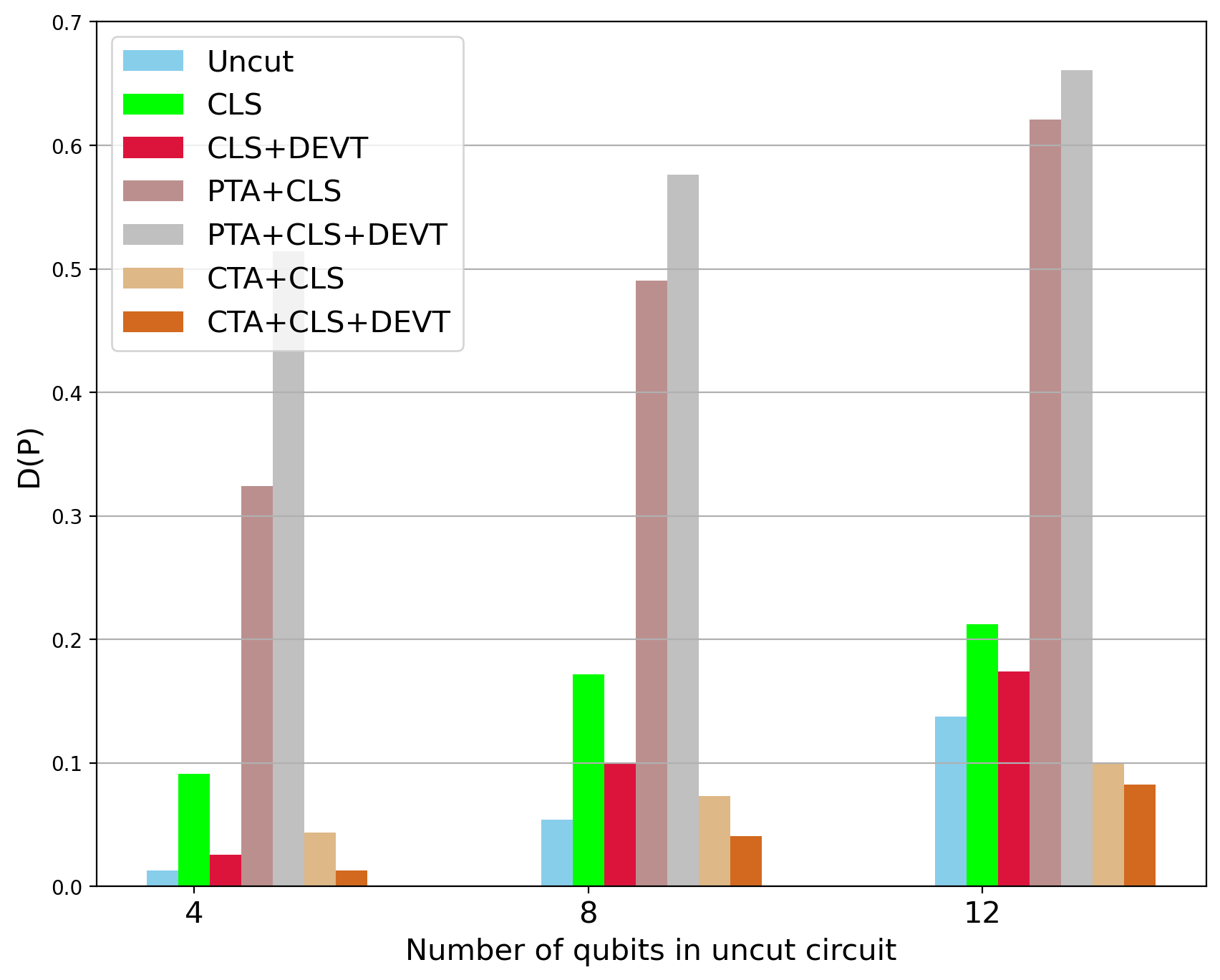}
    }
    \subfigure[Coherent error with $\Delta \theta = \frac{\pi}{32}$]{
        \includegraphics[scale=0.35]{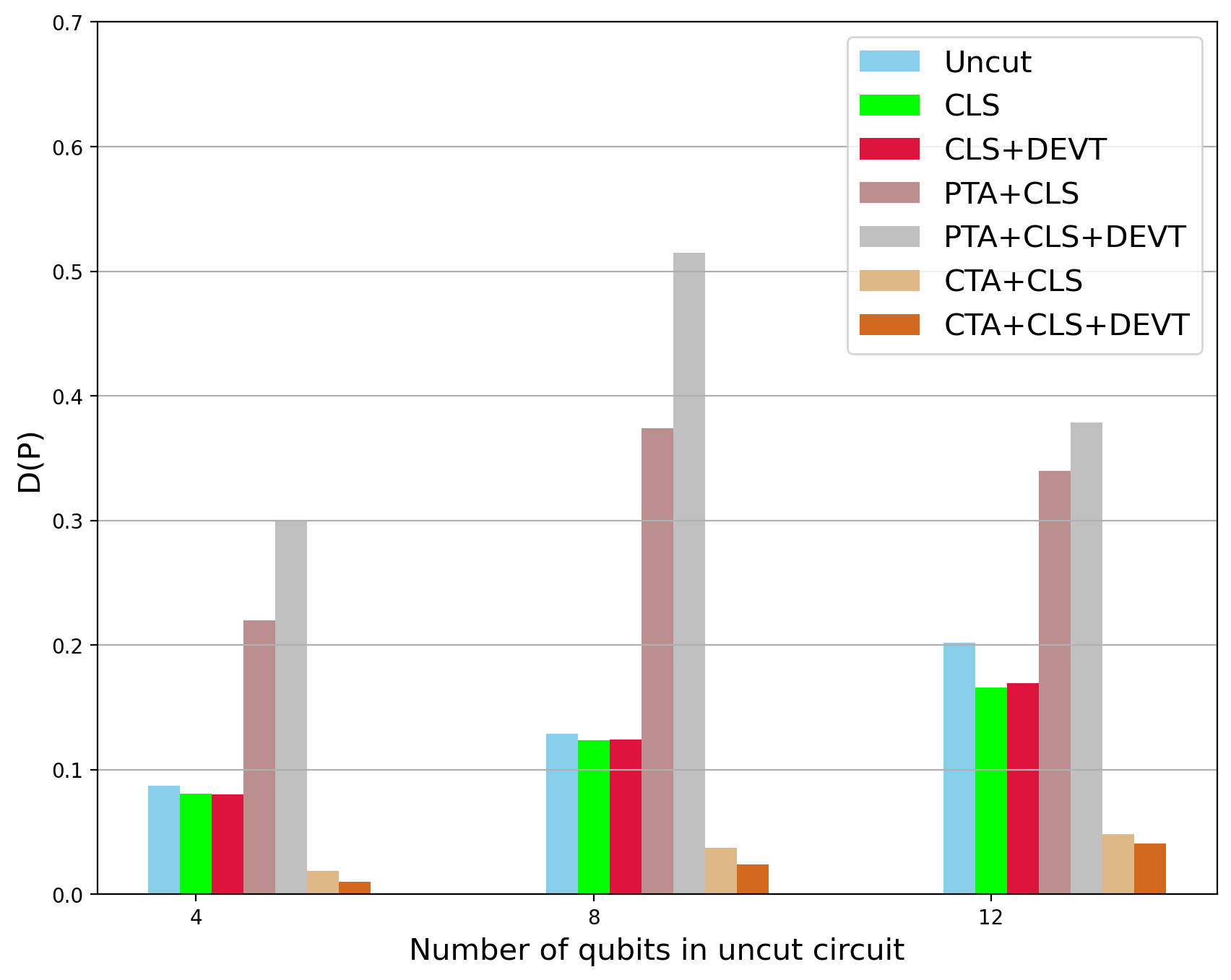}
    }
    \caption{Performance of error mitigated 2-fragment tomographic circuit cutting reconstruction using the Pauli-twirled approximation (PTA) and Clifford twirled approximation (CTA) of 2-qubit tensor product amplitude damping noise with $\gamma=0.01$ (a) and coherent noise with $\Delta\theta=\frac{\pi}{32}$ (b). Cut circuit reconstruction was compared using linear inversion (LIN), constrained least-squares (CLS) tomography fitters, both with and without dominant eigenvalue truncation (DEVT) mitigation and PTA or CTA. Direct measurement of the original circuit (uncut) is shown for comparison.}
    \label{fig:twirling}
\end{figure*}

Next, we consider a coherent error of the form 
\begin{equation}
    U_{err} = exp(-i \Delta\theta H_{\scriptsize{CNOT}}),
    \label{eq:coh_error}
\end{equation}
where $H_{\scriptsize{CNOT}} = log(U_{CNOT})/(-i)$ is the generator of a CNOT as a rotation gate, which is an approximate model of coherent errors due to imperfect gate calibration. The results for values of $\Delta\theta = \frac{\pi}{64}$ and $\Delta\theta = \frac{\pi}{32}$ are shown in \cref{fig:coherent} (a) and \cref{fig:coherent} (b) respectively. In this case we find that DEVT provides essentially no improvement to regular tomography fitting, though importantly we see that it also does not make the circuit cutting reconstruction significantly worse. One additional observation is that for the largest fragment size the cut circuit performs better than the uncut circuit, likely due to there being less total gates for the coherent error to accumulate in a single fragment. Moreover, for both of these noise models, we have not considered measurement error, and therefore, MEMCLS essentially becomes equivalent to CLS.

For the case of amplitude damping, the gate errors will also affect the conditioning qubit measurement outcomes, which could amplify error in the tomographic reconstruction. This is because, for such a noise model, not only the individual circuit fragments are expected to be erroneous, but also the conditional qubits are more accumulated towards some value, thus making the entire reconstruction severely faulty. Similarly, for coherent noise, the noisy density matrix is expected to deviate significantly from the ideal one, so that the overlap of the largest eigenvalue with the ideal state lowers. DEVT is not expected to generate fruitful result in such scenarios \cite{koczor2021dominant}. However, our numerical results show that DEVT still achieves a better performance than simple circuit cutting even under such scenarios. Therefore, it seems safe to deduce that if circuit cutting is used in such noisy scenarios, it is still not detrimental to apply DEVT.

\subsection{DEVT with Twirled Noise}\label{sec:twirled}

We note that Pauli unitary 2-qubit gate errors on Clifford gates which, such as the CNOT gate used in our simulations, can be converted to a Pauli channel mixed-unitary noise model via application of Pauli-twirling with negligible overall cost to the gate count of the circuit~\cite{wallman2016rc}. In the case of an amplitude damping channel Pauli twirling will result in a biased Pauli channel with $p_x = p_y = \frac{\gamma}{4}$ and $p_z = \frac{(1-\sqrt{1-\gamma})^2}{4}$, which in terms of the Pauli noise model considered in \cref{sec:gate-results} corresponds to a negative bias parameter, i.e., the channel has a higher probability of Pauli X or Y error than that of Pauli Z error. For $\gamma = 0.01$, $p_x = p_y = 0.0025$, and $p_z = 6 \times 10^{-6}$. As shown in \cref{fig:pauli}, DEVT will be expected to perform best when the resulting Pauli channel is closer to a combination of 1 or more depolarizing channels on any of the collections of subsystems; in other words, when the bias is close to 0.

\cref{fig:twirling} shows the results for simulating both the Pauli-twirled approximation (PTA), and Clifford-twirled approximation (CTA) to the amplitude damping and coherent error noise models from \cref{sec:non-pauli}. Here we observe that applying DEVT to the biased Pauli channel resulting from PTA of these noise models leads to a significantly worse result as compared to the un-twirled noise models. In \cref{app:bias} we discuss the performance of PTA in detail and provide some analytical reason for such an observation. These results indicate that in the context of circuit cutting that Pauli twirling should only employed if the resulting noise is not highly biased and close to a depolarizing channel.

\begin{figure*}
    \centering
    \subfigure[Noiseless data where the uncut circuit contains 4 qubits]{
        \includegraphics[scale=0.35]{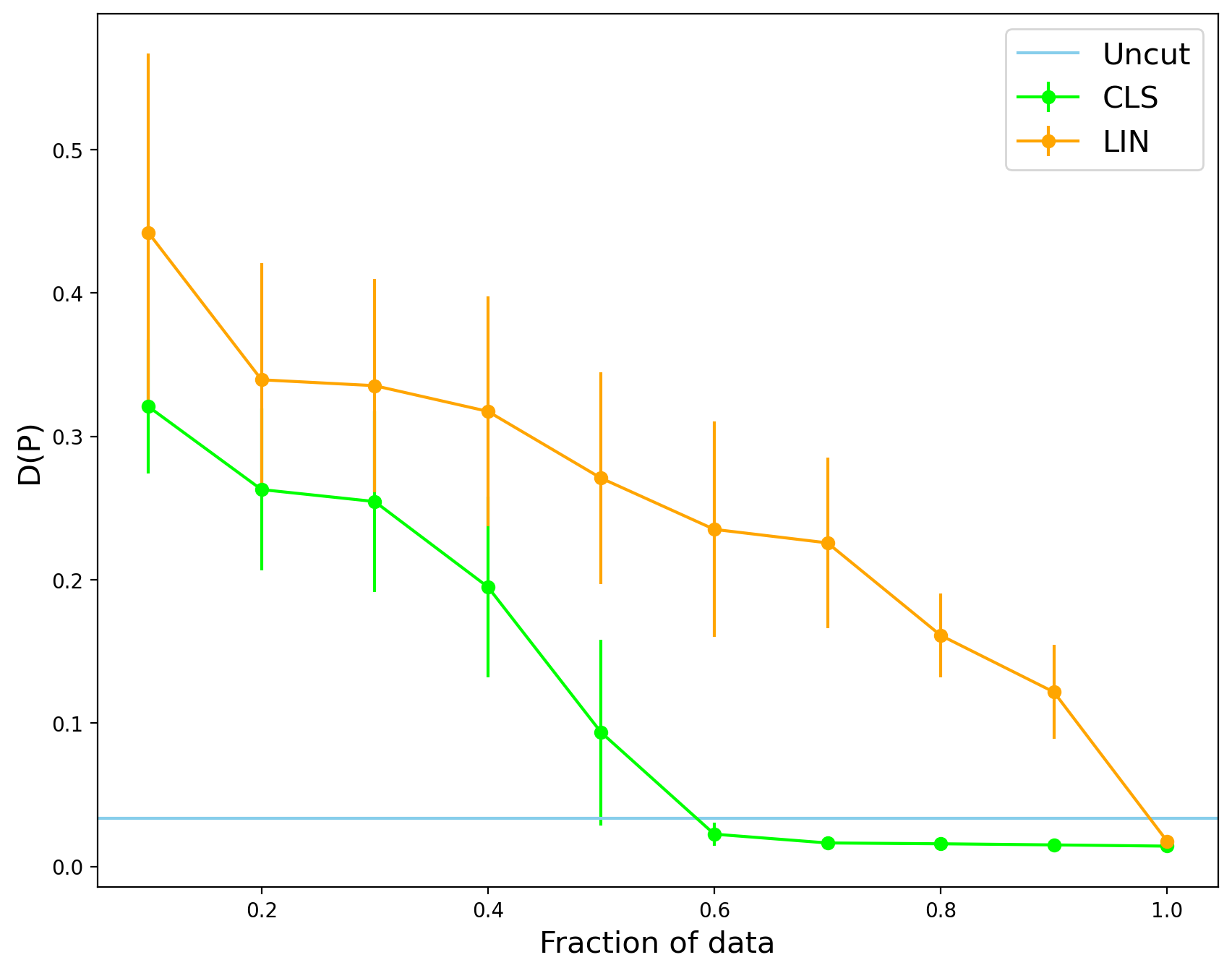}
    }\label{fig:pt_noiseless}
    \subfigure[Noisy data where the uncut circuit contains 4 qubits]{
        \includegraphics[scale=0.35]{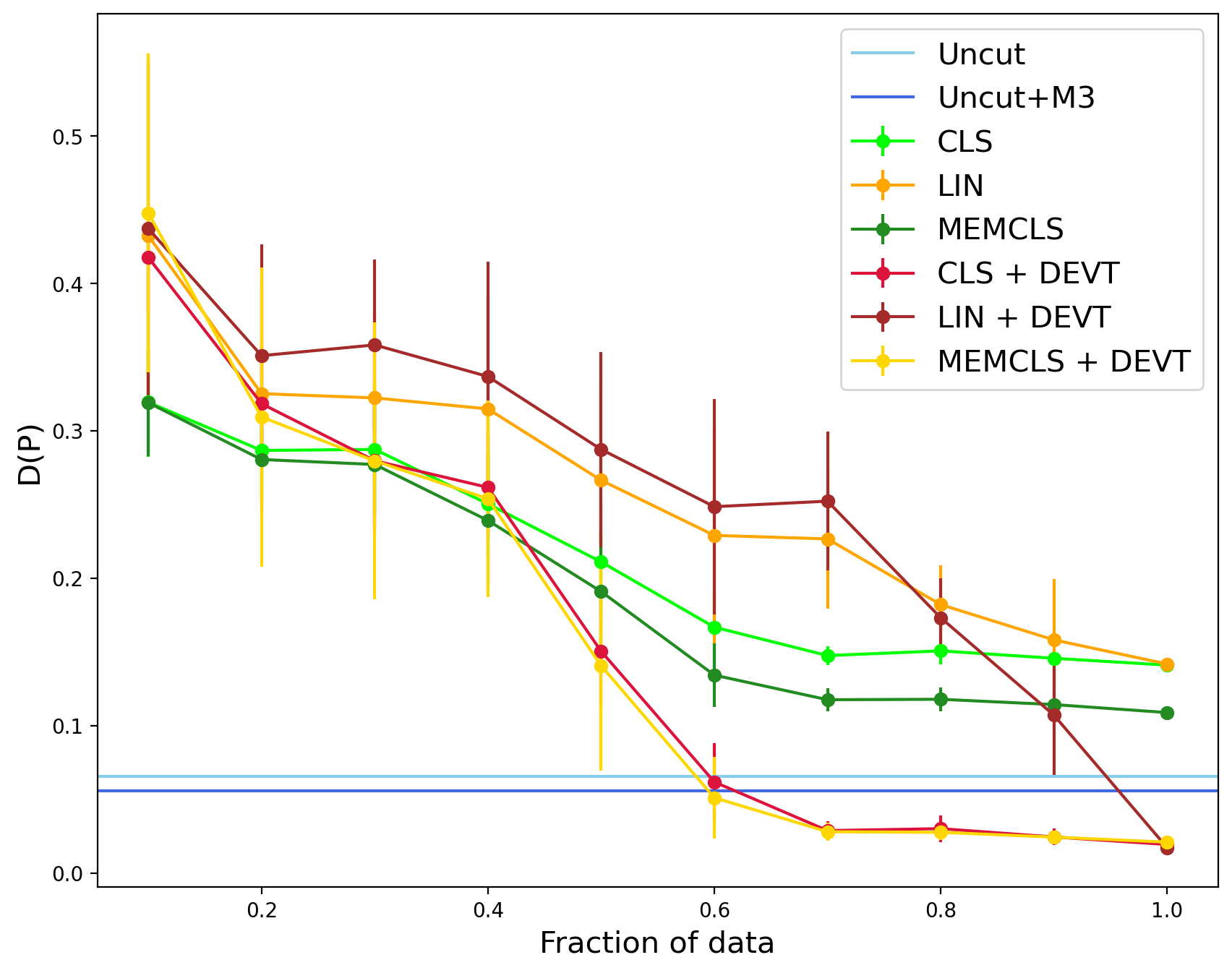}
    }\label{fig:pt_noise_2qbt}
    \subfigure[Noisy data where the uncut circuit contains 8 qubits]{
        \includegraphics[scale=0.35]{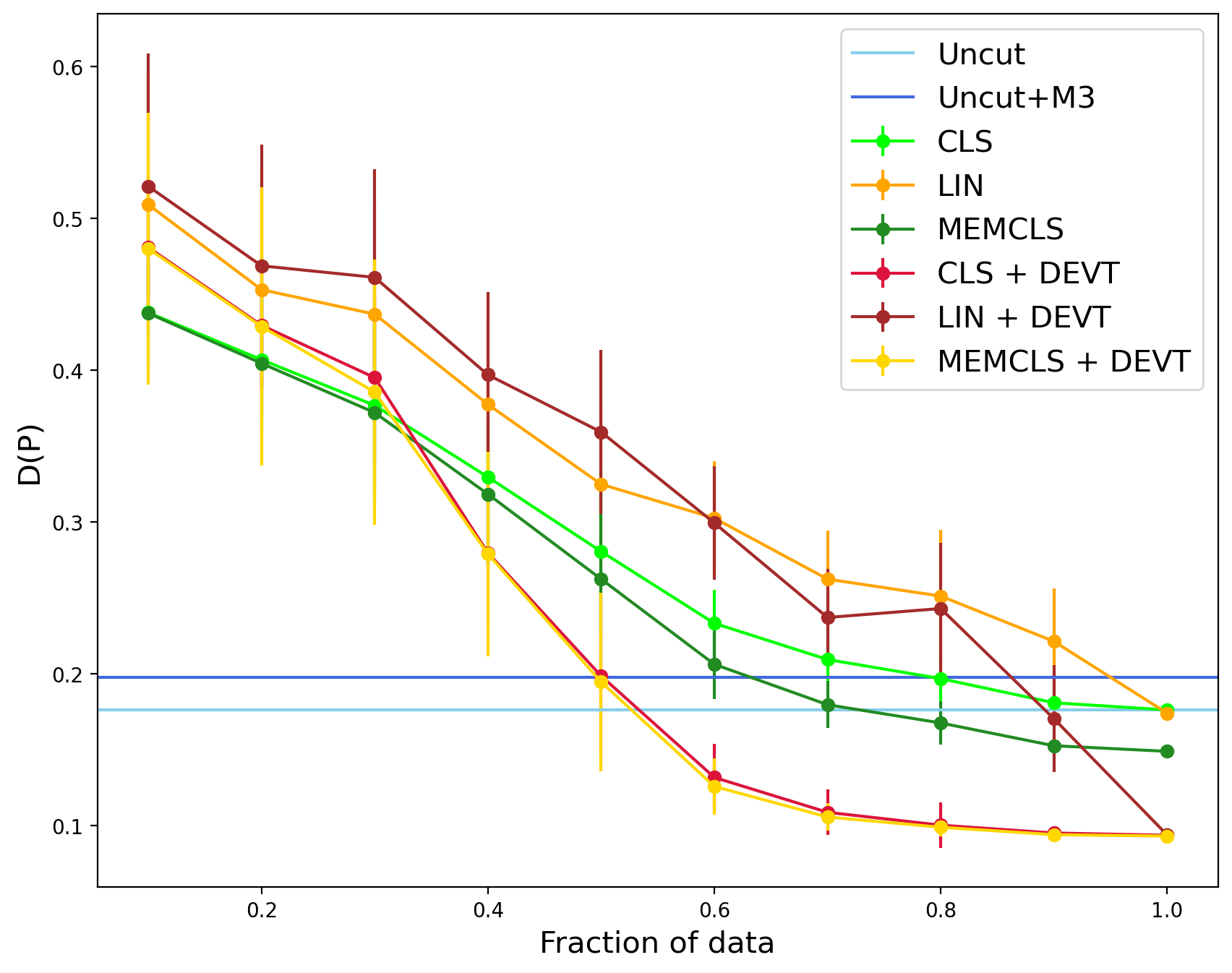}
    }\label{fig:pt_noise_4qbt}
    \subfigure[Noisy data where the uncut circuit contains 12 qubits]{
        \includegraphics[scale=0.35]{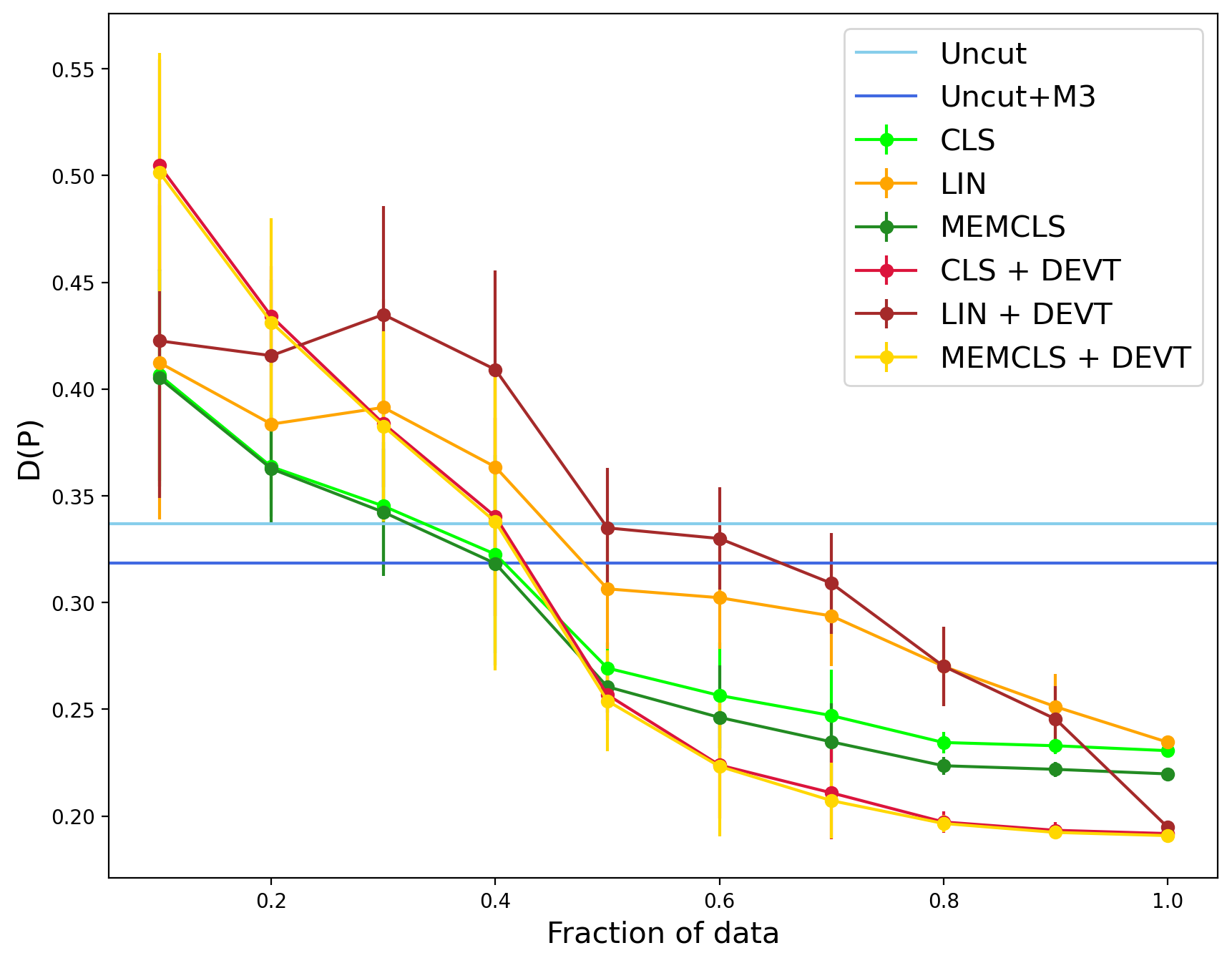}
    }\label{fig:pt_noise_6qbt}
    \caption{Performance of error mitigated 2-fragment tomographic circuit cutting reconstruction using partial tomography data. Data is a sampled as a subset of full data from \cref{fig:depol-c}, and averaged over 10 samples per data point. The noise model is a 2-qubit depolarizing gate noise with $p_{\scriptsize{depol}}=0.01$, local symmetric readout error with $p_{\scriptsize{meas}} = 0.05$, and single qubit gate depolarizing error of $p_1 = 10^{-4}$. Cut circuit reconstruction was compared using linear inversion (LIN), constrained least-squares (CLS) tomography fitters, readout error mitigated CLS (MEMCLS) both with and without dominant eigenvalue truncation (DEVT) mitigation. The original circuit (uncut) was measured with and without M3 readout error mitigation for comparison.}
    \label{fig:partial_tomo}
\end{figure*}

For the CTA, which converts an arbitrary noise channel into a depolarizing channel with the same average gate fidelity~\cite{magesan2012pra}, which is ideal for DEVT we observe a reduction in error as expected for the depolarizing noise model in \cref{sec:gate-results}. However, even when performing a tensor product of 1-qubit Clifford twirls, unlike with Pauli twirling, compiling these twirled gates into an arbitrary circuit can drastically increase the required number of 2-qubit gates, which makes it impractical for use when 2-qubit gates are the dominant error source.

\subsection{Circuit Cutting with Partial Data}
\label{sec:partial}

A $k$-qubit conditional tomography experiment using the standard Pauli basis and all measurement outcomes requires the execution of $12^k$ quantum circuits from the $4^k$ preparation states and $3^k$ measurement bases respectively. This means that full tomography is typically only practical for 2-3 qubit fragments in the process tomography case, or up 5-6 qubits for state tomography fragments. For larger number of qubits the classical post-processing required for linear inversion tomography can be significantly faster than for conditional least-squared tomography, which in the basic implementation of linear-least squares requires storing the full basis matrix of all vectorized basis elements. A natural question is whether partial tomography techniques are suitable for circuit cutting to reduce the number of experiments that need to be run. While there are many proposals for more scalable tomography fitters, we will investigate using the two standard fitters for linear inversion and constrained least squares with partial data.

Using the same experiment data as in \cref{sec:gate-results} with a 2-qubit depolarizing gate error of $p_2=0.01$, 1-qubit depolarizing gate error of $p_1 = 10^{-4}$ and symmetric readout error of $p_{\scriptsize{meas}}=0.05$ we perform tomographic reconstruction using a randomly sampled fraction $f$ of the full tomographic data ranging from $f=0.1$ to $f=1$, where the 100\% case corresponds to the previously presented results, the results of which are shown in \cref{fig:partial_tomo}. For each data point we have taken the average of $10$ trials for both the LIN and CLS to average over random basis selection. \cref{fig:partial_tomo} shows that when using partial data the CLS fitter greatly out performs linear inversion. This is to be expected as it has been shown that the positive constraints added to linear least squares is equivalent to compressed sensing tomography~\cite{kalev2015}.

In the absence of noise, \cref{fig:partial_tomo} (a) shows that CLS tomography performs at its maximum value when $f \simeq 0.6$. The standard deviation also drops to $\sim 0$ as the value of trace distance saturates. However, for LIN, the trace distance lowers almost linearly with increasing fraction of data, and attains its optimal trace distance only when the entire data set (i.e., $f = 1$) is used. This holds true in the presence of depolarizing gate and symmetric readout noise in \cref{fig:partial_tomo} (b), (c) and (d) respectively, though when noise is included the fraction of data required to match the full data case also increases. This is to be expected as more noisy states and channels are higher rank and will be less suitable with compressed sensing. Furthermore, as the number of qubits in the circuit inceases, so does the effect of noise, and the saturation point shifts towards higher values of $f$. In all cases with measurement error we find MEMCLS has slightly superior performance to CLS for all fractions of data, with the improvement increasing slightly the the fraction of data used.

When including DEVT mitigation we find that applying DEVT performs markedly better with partial data when using CLS and MEMCLS than with LIN. With both CLS and MEMCLS, DEVT provides a noticeable improvement when using $f>0.5$, i.e., $50\%$ of the data, with the exact value changing slightly with the fragments size, while for smaller fractions of data it increases the reconstruction error. For LIN, however, DEVT is only beneficial for $f>0.8$. 

In conclusion, if partial tomography measurement is used to reduce the overall number of experiments required for circuit cutting reconstruction, then there is a noticeable difference between CLS and LIN fitters, both with and without DEVT, and CLS should be strongly preferred over LIN. Furthermore, DEVT is still beneficial to reduce the overall error in the circut cutting reconstruction with partial data using $f>0.5$ of the data.

\section{Discussion}
\label{sec:discussion}

In this work we have explored how an error mitigated tomography approach to circuit cutting can improve the overall performance of evaluating quantum circuit outputs in the presence of gate and measurement noise. This builds on the previous work in \cite{perlin2021quantum} which showed advantages of using tomography over the original circuit cutting method \cite{peng2020simulating} in noiseless ideal simulations where only errors due to measurement sampling statistics were included. Across all simulations we observed that in the presence of gate noise, the circuit cutting reconstructed probabilities exhibit a greater sensitivity to the gate noise strength than the uncut circuit. This was true across all noise models considered, and emphasises the importance of error mitigation techniques that can be applied to circuit cutting. Our simulations demonstrated that in the presence of symmetric readout error measurement noise and certain forms of gate noise we can greatly improve tomographic circuit cutting estimates by applying DEVT during tomography reconstruction. For non-symmetric readout noise this can be made to look symmetric by Pauli twirling of the measurements to randomly flip the expected bit outcomes and then correcting in post-processing~\cite{ewout2022pra}. The form of the gate noise is important for DEVT to be effective, and in particular uniform noise close to a depolarizing channel is most effective with DEVT. While DEVT was found to provide little advantage for amplitude damping and coherent noise, it did not significantly increase the error in the reconstruction. One important result is that DEVT was shown to perform very poorly for highly biased Pauli noise, and could dramatically increase the error in the reconstruction. This is an important consideration to keep in mind if techniques such as Pauli-twirling are used to convert gate noises to Pauli channels. One possible way that this might be circumvented is to use probabilistic error amplification techniques, such as used in \cite{ying2017prx}, to make a highly biased gate noise more uniform, and hence more amenable to DEVT mitigation. 

Another important consideration when performing tomographic circuit cutting is the choice of tomography fitting procedure. The required time for post-processing can be a significant factor in applying circuit cutting to problems that have more than a handful of circuit fragments to be evaluated. However, since the tomographic reconstruction of each fragment is independent from other fragments this can be easily parallelized on classical computing resources.  We compared two of the most commonly used full tomography fitters, namely linear inversion and constrained least-squares optimization. In typical tomography applications the main trade-off between these two fitters is that linear inversion fitting is significantly faster, while constrained least squares is more accurate, especially when including readout error mitigation via noisy basis elements in the fitting or only using partial tomography data. In the context of circuit cutting with DEVT mitigation we found that linear inversion was comparable to constrained least squares when full tomographic data was available, and in particular DEVT was effective at mitigating the effect of measurement errors in tomography without requiring the specialized measurement-error mitigated conditional tomographic fitter we proposed. This can allow for significantly faster tomographic post-processing.

If circuit cutting is to be considered for applications requiring a number of cut qubits that is not realistic for obtaining full tomographic data then partial tomographic techniques will be required for the reconstruction. We found that in this case constrained least squares method performed significantly better than linear inversion, both with and without DEVT, since this method exhibits properties of compressed sensing. This also indicates that estimation techniques such as classical shadow tomography is most likely not an important consideration for circuit cutting since techniques such as shadow estimation~\cite{huang2020}, which are equivalent to linear inversion partial tomography in the Pauli basis as considered here, should not be expected to provide a benefit for circuit cutting problems over constrained tomography fitting methods, and further more are not suitable for use with DEVT as a mitigation method to improve their performance with partial data.

\section*{Acknowledgement}

The authors thank Ali Javadi-Abhari, Zlatko Minev, Alireza Seif, and Bryce Fuller for insightful discussion while working on this project. Ritajit Majumdar would like to acknowledge Fulbright Nehru Doctoral Research Fellowship for supporting this research, and United States-India Educational Foundation (USIEF) for awarding him with the same. He would also like to acknowledge IBM Thomas J Watson Research Center for hosting him during the research period. This research used resources of the National Energy Research Scientific Computing Center, a DOE Office of Science User Facility supported by the Office of Science of the U.S. Department of Energy under Contract No. DE-AC02-05CH11231 using NERSC award DDR-ERCAP0022238.

\bibliography{main}

\appendix

\section{DEVT with Measurement Errors}\label{app:devt-meas}

Our results in \cref{fig:mem} already establish numerically that DEVT mitigates measurement error better than MEMCLS, and padding MEMCLS with DEVT does not improve the result any further. In other words, DEVT is self sufficient for measurement error mitigation. Here we analytically show for linear inversion method of tomography that tomography with measurement error results in a noisy density matrix (or Choi matrix) of the form $\mathcal{E}(\rho) = (1-p)\rho + p\rho_{err}$. Therefore, DEVT alone is sufficient to mitigate the effect of measurement error.

Let us suppose that $\{\Pi_j\}$ is a tomographically complete basis, with each $\Pi_j$ being a projector. However, due to measurement error, the projectors are replaced by POVMs of the form $\tilde{\Pi_j} = (1-p) \Pi_j + p \Pi_j'$, $p$ being the probability of measurement error, and $\Pi_j'$ is the linear combination of one or more unwanted projectors forming the POVM. If $\rho$ be the state which is being measured, then the probability of success after measuring $\tilde{\Pi_j}$ is
\begin{eqnarray*}
\tilde{p_j} &=& Tr[\tilde{\Pi_j}\rho] \\
&=& (1-p) Tr[\Pi_j \rho] + p Tr[\Pi_j' \rho]\\
&=& (1-p) p_j + p p_j'
\end{eqnarray*}

The recreation of the state is carried out by creating the dual basis $\dket{D_j} = (\sum_j \dketdbra{\Pi_j}{\Pi_j})^{-1}\dket{\Pi_j}$~\cite{dariano2000pla}. Since, it is not expected that the exact form of POVM due to noise is known, we can assume that the dual basis remains the same irrespective of the noise. Therefore, the recreated state $\tilde{\rho}$
\begin{eqnarray*}
\tilde{\rho} &=& \sum_j \tilde{p_j} D_j \\
&=& (1-p) \sum_j p_j D_j + p \sum_j p_j' D_j \\
&=& (1-p) \rho + p \rho_{err}
\end{eqnarray*}

Therefore, as long as the largest eigenvalue of $\tilde{\rho}$ has a significant overlap with $\rho$, DEVT is sufficient to reestablish the error-free state $\rho$ from the state $\tilde{\rho}$ created due to measurement error. In other words, DEVT alone is sufficient to mitigate measurement errors.

\section{DEVT with depolarizing noise}\label{app:devt-depol}

Effect of depolarization error on a quantum state $\rho$ is denoted as in \cref{eq:depol_channel}. Note that the effect of depolarization noise model is readily similar to the required effect of noise for applying DEVT, as shown in \cref{eq:req} \cite{koczor2021dominant}. A depolarization channel remains depolarization even after $m \geq 1$ layers of gates. A single qubit state in a depolarization channel, after $m$ layers of gates, has the form $\rho_{noisy} = (1-p)^m \rho + [1-(1-p)^m]\frac{\mathbb{I}}{2}$, where, for simplicity, $p$ is assumed to be the probability of error for each gate. Therefore, for a $n$-qubit circuit with depth $m$, the effective erroneous state can be represented as
\begin{eqnarray}
    \label{eq:noisy_n_qbt}
    \rho_{noisy}^n &=& \otimes_{i=1}^n \rho_i^{noisy} \\ \nonumber
    &=& \otimes_{i=1}^n (1-p)^m\rho_i + [1-(1-p)^m]\frac{\mathbb{I}}{2} \\ \nonumber
    &=& (1-p)^{n.m} \rho_{out}^n + \rho_{err}.
\end{eqnarray}

Note that $\rho_{err}$ is a summation of multiple density matrices. It is not possible to ascertain the largest eigenvalue of $\rho_{err}$ without explicit information of $\rho_i$, $\forall$ $i$. Hence, the only consideration possible is the worst case scenario that the largest eigenvalue of $\rho_{err} \leq 1$. Therefore, putting $\delta \leq (\frac{1}{(1-p)^{n.m}}-1)$, we obtain an upper bound of the coherent mismatch $c$ for depolarization noise model.

\begin{eqnarray}
    \label{eq:ub_depol}
    c & \leq & \frac{\delta^2}{4} = \frac{1}{4}(\frac{1}{(1-p)^{n.m}}-1)^2 \\ \nonumber
    &=& \frac{1}{4} [\frac{1-(1- n.m.p + \mathcal{O}(n^2 m^2 p^2))}{1- n.m.p + \mathcal{O}(n^2 m^2 p^2)}]^2 \\ \nonumber
    & \approx & \frac{1}{4}[\frac{n.m.p}{1-n.m.p}]^2 = \mathcal{O}((n.m.p)^2)
\end{eqnarray}

\section{DEVT with Pauli noise}\label{app:devt-pauli}

In this subsection, we briefly touch upon the seemingly worse performance of DEVT for pauli noise as opposed to that of depolarization noise. The evolution of a density matrix under pauli noise is shown in \cref{eq:pauli_noise}, which conform to the form required for DEVT as in \cref{eq:req}. However, consider a circuit which applies $k$ layers of gates on the input density matrix $\rho_{in}$. The ideal output density matrix $\rho_{out}$ is, thus, given by
\begin{equation*}
    \rho_{out} = G_k G_{k-1} \hdots G_1 \rho_{in} G_1^{\dagger} \hdots G_{k-1}^{\dagger} G_k^{\dagger}
\end{equation*}

We consider the noisy implementation of each gate layer $G_i$ to be $G_i' = P_i G_i$, where $P_i$ consists of one or more pauli errors. Under the action of such a noisy channel, the noisy output density matrix becomes
\begin{center}
    $\rho_{out}^{noisy} = \Pi_{i=1}^{k} P_i G_i \rho_{in} G_i^{\dagger} P_i$
\end{center}

Since, in general, each $G_i$ does not necessarily consist of Clifford gates only, we have the following scenario:
\begin{center}
    $\rho_{out}^{noisy} = P_k P_{k-1} G_k G_{k-1} \Pi_{i=i}^{k-2} (P_i G_i \rho_{in} G_i^{\dagger} P_i) G_{k-1}^{\dagger} G_k^{\dagger} P_{k-1} P_k + [P_{k-1},G_k]$
\end{center}

In other words, the noisy output density matrix takes the eventual form
\begin{equation}
    \label{eq:pauli}
    \rho_{out}^{noisy} = \Pi_{i=1}^k P_i (\Pi_{j=1}^k G_j \rho_{in} G_j^{\dagger}) P_i + comm
\end{equation}
where \emph{comm} denotes all the commutator terms. From \cref{eq:pauli} we note that pauli noise results in some commutator terms along with the required form of \cref{eq:req}. Therefore, the eventual form of the noisy density matrix deviates from \cref{eq:req}, resulting in a poorer performance of DEVT for this noise model. It may be interesting to study the performance bounds of DEVT for certain circuits of importance with pauli noise, especially if the number of non-Clifford gates and their distribution is known beforehand. However, we postpone that study for the future.

\subsection{Dominant eigenvalue for biased noise channel}
\label{app:bias}
Here we dive deeper into the DEVT method for Pauli channel of the form of \cref{eq:pauli_noise}. For an arbitrary pure state $\rho = \begin{pmatrix}
|\alpha|^2 & \alpha^* \beta \\
\alpha \beta^* & |\beta|^2
\end{pmatrix}$, where $\alpha, \beta \in \mathbb{C}$, $\rho \geq 0$, and $Tr\{\rho\} = 1$, the eigenvalue is $1$ with eigenvector $\ket{\psi} = \alpha\ket{0} + \beta\ket{1}$. Now, for a Pauli channel of the form of \cref{eq:pauli_noise}, the difference between the ideal density matrix $\rho$, and the noisy density matrix $\mathcal{E}_{\scriptsize{pauli}}(\rho)$ takes the form of \cref{eq:density_pauli}.
\begin{equation}
    \label{eq:density_pauli}
    \rho_{\scriptsize{diff}} = \rho - \mathcal{E}_{\scriptsize{pauli}}(\rho) = \begin{pmatrix}
    2p(|\alpha|^2-|\beta|^2) & (4p+2b)\alpha\beta^{*}\\
    (4p+2b)\alpha^{*}\beta & 2p(|\beta|^2-|\alpha|^2
    \end{pmatrix}
\end{equation}

Note that for successful performance of DEVT, we require the dominant eigenvector of the erroneous density matrix $\mathcal{E}_{\scriptsize{pauli}}(\rho)$ to share a large overlap with the ideal pure state $\rho$. In other words, we require the dominant eigenvalue of $\mathcal{E}_{\scriptsize{pauli}}(\rho) \rightarrow 0$. Through the usual method to calculate eigenvalue, we find the eigenvalues of $\rho_{\scriptsize{diff}}$ to be $\lambda = \pm \sqrt{p^2 + 4pb |\alpha|^2 |\beta|^2 + b^2 |\alpha|^2 |\beta|^2}$. Since, we are interested only in the dominant eigenvalue of $\rho_{\scriptsize{diff}}$, we look a bit closely to $\lambda_{\scriptsize{dominant}} = \sqrt{p^2 + 4pb |\alpha|^2 |\beta|^2 + b^2 |\alpha|^2 |\beta|^2}$.

When the bias $b = 0$, then $\lambda_{\scriptsize{dominant}} = p$. In other words, for a depolarizing channel, the dominant eigenvalue of $\rho_{\scriptsize{diff}}$ is simply the probability of error. Note that since $p$, $|\alpha|^2$ and $|\beta|^2$ are non-negative, if $b > 0$, then $\lambda_{\scriptsize{dominant}} > p$, i.e., the dominant eigenvalue of $\rho_{\scriptsize{diff}}$ increases with increasing bias. Therefore, the performance of DEVT for a Pauli channel with positive bias is expected to be worse than depolarizing channel.

If $b < 0$, as for the amplitude damping channel, then it may be possible to have $\lambda_{\scriptsize{dominant}} < p$ if
\begin{eqnarray}
\label{eq:pta_cond}
    4pb |\alpha|^2 |\beta|^2 + b^2 |\alpha|^2 |\beta|^2 &<& 0 \nonumber \\
    \Rightarrow p &>& -\frac{b}{4}
\end{eqnarray}

In the amplitude damping channel, we considered $\gamma \in \{0.001, 0.01\}$, and showed that Pauli twirl approximation of such a channel leads to an effective Pauli channel with $p_x = p_y = \frac{\gamma}{4}$ and $p_z = \frac{(1-\sqrt{1-\gamma})^2}{4}$. In \cref{tab:pta} we show the value of the bias $b$ for the two values of $\gamma$, and the required value of $p$ to obtain $\lambda_{\scriptsize{dominant}} < p$ according to \cref{eq:pta_cond}.

\begin{table}[htb!]
    \centering
    \caption{Required value of error probability $p$ to satisfy \cref{eq:pta_cond}}
    \resizebox{85mm}{!}{
    \begin{tabular}{|c|c|c|c|c|c|}
    \hline
        $\gamma$ & $p_x$ & $p_y$ & $p_z$ & bias $b$ & \scriptsize{Req$^d$ $p$} \\
        \hline
        $0.001$ & $2.5 \times 10^{-4}$ & $2.5 \times 10^{-4}$ & $6 \times 10^{-8}$ & $-0.9999$ & $0.249975$ \\
        \hline
        $0.01$ & $2.5 \times 10^{-3}$ & $2.5 \times 10^{-3}$ & $6 \times 10^{-6}$ & $-0.999$ & $0.24975$\\
        \hline
    \end{tabular}}
    \label{tab:pta}
\end{table}

The values from \cref{tab:pta} shows that unrealistically high value of error probability is required for the bias under consideration for the $\lambda_{\scriptsize{dominant}}$ to be lower than $p$. In other words, it can be safely concluded that bias in the channel leads to a poorer performance of DEVT than a depolarization channel in every scenario.

\end{document}